\documentclass[12pt, a4paper, oneside]{article}
\usepackage{comment}
\usepackage[utf8]{inputenc} 
\usepackage[hidelinks]{hyperref}
\usepackage[left=2cm,top=2 cm,right=2cm,bottom=2cm,bindingoffset=0.5cm]{geometry} 
\usepackage{amsmath} 
\usepackage{amssymb} 
\usepackage{subcaption}
\usepackage{float} 
\usepackage{enumerate} 
\usepackage{amsthm} 
\usepackage{mathrsfs} 
\usepackage{braket}
\usepackage{mathtools}
\usepackage{indentfirst}
\usepackage{cases}
\usepackage{authblk}
\usepackage{cite}
\usepackage{color}

\renewcommand{\thesection}{\arabic{section}}
\renewcommand{\theequation}{\thesection.\arabic{equation}}

\newcommand{\be}{\begin{equation}}
\newcommand{\ee}{\end{equation}}
\newcommand{\bea}{\begin{eqnarray}}
\newcommand{\eea}{\end{eqnarray}}

\newcounter{orange}
\renewcommand{\theorange}{\alph{orange}}

\newcommand\tagg{\addtocounter{equation}
{1}\tag{\theequation}}

\begin{document}

\title{Phase space quantum mechanics of a free particle moving in a plane}

\author{Hubert Jóźwiak \footnote{E-mail ad\text{d}ress: hubert.jozwiak@dokt.p.lodz.pl  (corresponding author)}}
\author{ Jaromir Tosiek \footnote{E-mail ad\text{d}ress: jaromir.tosiek@p.lodz.pl}}
\affil[1]{Institute of Physics, Lodz University of Technology, ul. W\'olcza\'nska 217/221, 93-005 \L\'od\'z, Poland}
\date{\today} 
 \maketitle

 \begin{abstract}
 The eigenstates of a free quantum particle propagating in a plane are derived in the context of  phase space quantum mechanics. Two possibilities are analysed. First, the case of a particle with fixed energy and angular momentum is considered. A special choice of coordinates on the four-dimensional phase space, suitable for representing the eigenstates of the particle under consideration, is presented. A phase-space counterpart of the product of operators, known as 
 the Moyal product, is derived in these coordinates. The eigenvalue equations  are solved, and their physically acceptable solutions, called Wigner functions, are identified. Second, a particle with fixed components of the Cartesian momentum is considered. A relationship is found between the Wigner function of the particle with fixed Cartesian momentum components and some functions characterising the particle with fixed energy and angular momentum.

   \end{abstract}

\section{Introduction}
Phase space quantum mechanics  ~\cite{YK91, FS94, WS01, CZ05, tat, cd, lee, dit, SW07} is an interesting alternative to the Hilbert-space version of quantum theory. Its power becomes apparent when applied to systems with nontrivial configuration spaces, but it can also be fruitful in more elementary problems if non-Cartesian charts are used. Moreover, the semiclassical approximation arises naturally within this framework. 

For physical systems whose classical phase space is $\mathbb{R}^{2n}$, the quantum phase space coincides with the classical phase space, and quantum observables are represented by smooth real-valued functions on this space. By contrast, states are not identified with points in phase space, but with quasiprobability distributions known as Wigner functions. These functions are real-valued but not necessarily positive-valued.
Examples of  problems solved in this manner can be found in  \cite{IS1,IS2,IS3,IS4,IS5,IS6}.
 
  Moreover, the multiplication of quantum observables is no longer the standard pointwise multiplication `$\cdot$' of functions.  Instead, it is given by an associative, bilinear operation `$\star_{(\text{M})}$', known as the Moyal product. 
  Its general form is expressed by an integral formula, but in several cases one can apply its differential version. 
  
  Thus, the star-eigenvalue equations for observables 
  are usually  partial differential equations for  Wigner functions.  Such equations 
   may be difficult to solve. One of the standard methods for simplifying PDEs is a change of coordinates. In the case of quantum  star-eigenvalue equations, a transformation of coordinates can be performed with the use of the Fedosov construction~\cite{Fedosov1, Fedosov2}.  The Fedosov algorithm involves advanced differential geometry and is dedicated to calculations on curved phase spaces \cite{F1, F2, F3}. However, it can be adapted to
    flat spaces and in this case its application is straightforward.

In a previous work \cite{Tosiek_oscillator}, we presented a method for solving the $\star_{(\text{M})}$-eigenvalue problem for a 1D Hamiltonian. In that publication, we constructed the star-eigenvalue equation in a convenient coordinate system by deriving the  $\star_{(\text{M})}$-product using  only the Fedosov method.

In the current article, we extend that approach  to solve the star-eigenvalue problem for a free quantum particle moving in a plane, referred to as a 2D free particle. In contrast to classical physics, there are two types of 2D quantum free particles---particles with fixed Cartesian momentum components or with fixed energy and angular momentum. In Hilbert-space quantum mechanics the relationship between wave eigenfunctions of these two kinds of free particles is known as the Jacobi-Anger expansion.  We propose the phase-space counterpart of this expansion with the use of so-called cross-Wigner functions.

Calculating the eigenstates of a two-dimensional free quantum particle with specified Cartesian momentum components is straightforward. By contrast, deriving the eigenstates of a free particle characterised by its energy and angular momentum is considerably more demanding. For this reason, we choose a special set of Darboux coordinates $(T, \chi, H, L)$---i.e. coordinates in which the symplectic form has canonical form.
Then, employing the adapted Fedosov algorithm, we explicitly construct the $\star_{(\text{M})}$-product in the new chart. We formulate the $\star_{(\text{M})}$-eigenvalue problem for both the Hamiltonian $H$ and the angular momentum $L$ and solve two $\star_{(\text{M})}$-eigenvalue problems explicitly, obtaining a collection of Wigner functions $W_{Em}.$  Finally, we examine  which of the resulting functions represent  physically acceptable  Wigner functions of the 2D free particle of fixed energy $E$ and the magnetic quantum number $m$. The final part of our work addresses the vital question of how to decompose states into combinations of certain quasiprobability distributions. Namely, 
knowing the Wigner function $W_{p_{x0}p_{y0}}$ for the 2D free particle with the fixed components  of the Cartesian momentum $p_{x0}, p_{y0}$ and some cross-Wigner functions $W_{Emm'}$ for the 2D free particle with fixed energy $E$ and angular momenta $m\hbar, m'\hbar$, we find an expansion of $W_{p_{x0}p_{y0}}$ in the generalised linear combination of functions from the set $\{W_{E m m'}\}_{E>0;m,\,m'\in\mathbb{Z}}.$ This result illustrates the structure of the set of states as a linear space.

Wigner functions $W_{Em}$ have already been studied in the literature \cite{Schleich0, Schleich}. However, the cross-Wigner functions $W_{Emm'}$ and the phase-space representation of the Jacobi-Anger expansion constitute new results. Novelty is also seen in the fact that we 
obtain all results exclusively within the framework of phase-space quantum mechanics, which proves that this attempt to apply quantum physics is autonomous and consistent. Moreover, our considerations indicate the power of geometry, which enables us to choose preferred coordinates on the phase space.


\section{The Wigner functions of a 2D free particle}
\setcounter{equation}{0}
Consider a free nonrelativistic classical  particle of mass $M$ moving in $\mathbb{R}^2$ space. Its state can be uniquely determined by the measurement of position $\mathbf{r} = (x,\,y) \in \mathbb{R}^2$ and momentum $\mathbf{p}=(p_x,\,p_y) \in \mathbb{R}^2$. The collection of all pairs $(\mathbf{r},\,\mathbf{p})$ constitutes the space $\mathbb{R}^4$. However, 
the linear space $\mathbb{R}^4$ is not sufficient to model physical systems---e.g. to represent their time evolution. Thus, we need to add an extra structure represented by a tensor denoted as $\omega_{ij}$ and called the symplectic structure \cite{Arnold, Libermann}. In Cartesian coordinates $(x,y,p_x,p_y)$, the explicit form of this tensor is given by the first formula in \eqref{xx}.
 The pair $(\mathbb{R}^4,\,\omega)$ is  known as the symplectic manifold. 

The states of this classical particle are characterised by three independent constants of motion: the components of the Cartesian momentum $p_{x}, p_{y}$ and the angular momentum $L = x p_y - y p_x.$

The states of a free quantum particle on the plane $\mathbb{R}^2$ are  characterised either by the values of the energy, given by the Hamilton function $H$, and of the angular momentum $L$, or, alternatively, by components of the Cartesian momentum $p_x, p_y$. In contrast to classical mechanics, these two  kinds of quantum states are not equivalent---i.e. the quantum particle with defined components of the Cartesian momentum has an undetermined angular momentum, and vice versa.

The quantum phase space of this particle is---as in classical physics---the symplectic manifold $(\mathbb{R}^4,\, \omega).$ We need  to introduce the Moyal product on the symplectic manifold. This product is a counterpart of the composition of linear operators in a Hilbert space. The Moyal product $\star_{(\text{M})}$ 
of functions $f(\mathbf{r}, \, \mathbf{p})$ and $g(\mathbf{r}, \, \mathbf{p})$
is defined (compare \cite{Przanowski, funkcje}) in the Cartesian chart $(\mathbf{r},\,\mathbf{p}) = (x,\,y,\,p_x,\,p_y)$ as
\begin{align}
\notag
    &(f\star_{(\text{M})} g)(\mathbf{r},
    \, \mathbf{p}) := \frac{1}{(\pi \hbar)^4}\int_{\mathbb{R}^{8}}\text{d}\mathbf{r}'\text{d}\mathbf{p}'\text{d}\mathbf{r}''\text{d}\mathbf{p}'' \\
    &\times f(\mathbf{r}'
    , \mathbf{p}')g(\mathbf{r}''
    , \mathbf{p}'')\exp\Bigg(\frac{2i}{\hbar}[(\mathbf{r} - \mathbf{r}') (\mathbf{p} - \mathbf{p}'') - (\mathbf{r} - \mathbf{r}'')(\mathbf{p} - 
    \mathbf{p}')]\Bigg).
\end{align}
However, in several cases  one can apply  
 the differential formula
\begin{align}
\label{star-exp}
    (f \star_{(\text{M})} g)(\mathbf{r},\,\mathbf{p}) &:= f(\mathbf{r},\,\mathbf{p}) \; \text{exp}\left(\frac{i\hbar}{2} \left[\frac{\overleftarrow{\partial}}{\partial x}\frac{\overrightarrow{\partial}}{\partial p_x} - \frac{\overleftarrow{\partial}}{\partial p_x}\frac{\overrightarrow{\partial} }{\partial x}+ \frac{\overleftarrow{\partial}}{\partial y}\frac{\overrightarrow{\partial}} {\partial p_y}- \frac{\overleftarrow{\partial}}{\partial p_y}\frac{\overrightarrow{\partial}}{\partial y} \right]\right)g (\mathbf{r},\,\mathbf{p}).
\end{align}
The arrows indicate which function is differentiated. The symbol $\exp(\cdot)$ represents the series  $\sum_{j=1}^{\infty}\frac{(\cdot)^j}{j!}.$ Thus, we see that the Moyal product \eqref{star-exp} consists of derivatives of unbounded degree.

The Moyal bracket $\{\cdot,\,\cdot\}_{(\text{M})}$ 
of  functions $f(\mathbf{r}, \, \mathbf{p})$ and $g(\mathbf{r}, \, \mathbf{p})$,
being a counterpart of the commutator of respective operators, is defined as
\begin{equation}
    \{f,\,g\}_{(\text{M})}(\mathbf{r},\,\mathbf{p}) := \frac{1}{i\hbar}\big(f\star_{(\text{M})} g - g\star_{(\text{M})} f \big) (\mathbf{r},\,\mathbf{p}) .
\end{equation}

As already mentioned, the eigenstates of a 2D free quantum particle can be determined by two complete sets of observables $\{R_1,\,R_2\}$ commuting in the sense of the Moyal bracket: 
\begin{enumerate}
    \item[(i)] the energy and the angular momentum: $R_1:=H = \frac{p^2_x + p^2_y}{2M}$ and $ R_2:=L = xp_y - yp_x$,
     \item[(ii)] the components of the momentum: $ R_1:=p_x$ and $R_2:= p_y$.
\end{enumerate}
These two sets of observables are neither equivalent nor complementary, because \\
$
\{ p_x,\,L\}_{(\text{M})} \neq 0$ and $
\{ p_y,\,L\}_{(\text{M})} \neq 0$.

The Wigner function $W_{r_1r_2}(\mathbf{r},\,\mathbf{p})$ of the 2D free particle is a real function or a tempered distribution on $\mathbb{R}^4$ that satisfies
the system of $\star_{(\text{M})}$-eigenvalue equations
\setcounter{orange}{1}
\renewcommand{\theequation} {\arabic{section}.\arabic{equation}\theorange}
\begin{numcases}{}
\label{i1}
   (R_1\star_{(\text{M})} W_{r_1r_2})(\mathbf{r},\,\mathbf{p}) = r_1W_{r_1r_2}(\mathbf{r},\,\mathbf{p}), \\
     \addtocounter{orange}{1}
\addtocounter{equation}{-1}
    \label{i2}
   (R_2 \star_{(\text{M})} W_{r_1r_2})(\mathbf{r},\,\mathbf{p}) = r_2 W_{r_1r_2}(\mathbf{r},\,\mathbf{p}),
\end{numcases}
together with the commutativity conditions 
  \addtocounter{orange}{1}
\addtocounter{equation}{-1}
\begin{numcases}{}
   \label{i3}
    \{ R_1,\, W_{r_1r_2}\}_{(\text{M})}(\mathbf{r},\,\mathbf{p}) &= 0, \\
      \addtocounter{orange}{1}
\addtocounter{equation}{-1}
\label{i4} \{R_2,\,W_{r_1r_2}\}_{(\text{M})}(\mathbf{r},\,\mathbf{p}) &= 0,
\end{numcases}
\renewcommand{\theequation} {\arabic{section}.\arabic{equation}}
and the normalisation condition 
\begin{equation}
   \label{normalisation_def}(W_{r_1r_2}\star_{(\text{M})}W_{r'_1r'_2})(\mathbf{r},\,\mathbf{p}) \propto  \frac{1}{(2\pi\hbar)^2}f_{r_1r'_1}f_{r_2r'_2}W_{r_1r_2}(\mathbf{r},\,\mathbf{p}),
\end{equation}
where for $ i = 1,\,2$
\begin{align}
   f_{r_ir'_i} := \begin{cases}
       \delta(r_i - r'_i) &\text{, if the spectrum of $R_i$ is continuous}, \\
        \delta_{r_i r'_i} &\text{, if the spectrum of $R_i$ is discrete}.
    \end{cases}
\end{align}
By $r_1$ and $r_2$ we denote the eigenvalues of functions $R_1$ and $R_2$, respectively.

Moreover, the Wigner functions $W_{r_1r_2}(\mathbf{r},\,\mathbf{p})$ have to obey the condition of positivity of the marginal distributions
\begin{align}
\label{margin}
    \int_{\mathbb{R}^2} \text{d}\mathbf{r}W_{r_1r_2}(\mathbf{r},\,\mathbf{p}) &\geq 0,  &  \int_{\mathbb{R}^2} \text{d}\mathbf{p}W_{r_1r_2}(\mathbf{r},\,\mathbf{p}) &\geq 0.
\end{align}

We do not expect that any of the Wigner functions $W_{r_1r_2}(\mathbf{r},\,\mathbf{p})$  satisfy the condition \\
$
\int_{\mathbb{R}^4} \text{d}\mathbf{r}
\text{d}\mathbf{p}W_{r_1r_2}(\mathbf{r},\,\mathbf{p}) < \infty$,
as the particle under consideration is unbound.  Unbound states are rarely discussed in the literature on phase-space quantum mechanics  \cite{cont1,cont2,cont3, cont4}. 

\renewcommand{\theequation} {\arabic{section}.\arabic{equation}}

\section{Wigner functions of eigenstates of energy and angular momentum}
\setcounter{equation}{0}
Consider the 2D particle with fixed energy and angular momentum. One can see that $R_1 = H$ and $R_2 = L$ are qua\text{d}ratic functions of position and momentum. Hence, the Moyal brackets in Eqs. (\ref{i3})--(\ref{i4}) are proportional to the respective Poisson brackets. This means that if we find coordinates $T$ and $\chi$ such that $(x,\,y,\,p_x,p_y) \to (T,\,\chi,\,H,\,L)$ is the symplectic transformation \cite{Arnold}---i.e. a transformation that preserves the symplectic form---then the LHS of Eqs. (\ref{i3})--(\ref{i4}) takes its simplest form: $\{H,\,\cdot\}_{(\text{M})} = -\frac{\partial}{\partial T}(\cdot)$ and $\{L,\,\cdot\}_{(\text{M})} = -\frac{\partial}{\partial \chi}(\cdot)$. In order to derive the explicit expressions for $T$ and $\chi$, one considers preservation of the symplectic form $\omega$ under the aforementioned  transformation. We obtain
\begin{align}
\notag
   T &= \frac{M(xp_x + yp_y)}{p^2_x+p^2_y},   & \chi &= \text{atan2}(p_y,\,p_x), \\
   \label{new_coordinates}
     H &= \frac{p^2_x+p^2_y}{2M}, & L &= xp_y - yp_x.
\end{align}

The transformation (\ref{new_coordinates}) is well defined  only if $p_x^2+p_y^2>0$---i.e. it does not work for classical particles at rest.  Fortunately, the subset $p_x^2 + p^2_y = 0$ is a set of measure zero. Hence, it can be ignored from the quantum  point of view. One can check that the Jacobian of the transformation equals   
\begin{equation}
    \frac{\partial (T,\,\chi,\,H,\,L)}{\partial(x,\,y,\,p_x,\,p_y)} = 1.
\end{equation}

The classical interpretation of the variable $T$ is the  time of arrival for a particle of mass $M$  moving
with the velocity  $\frac{\mathbf{p}}{M}$
 from the origin $(0,\,0)$ to the point obtained by projecting the point ${\mathbf{r}}$ onto the direction of the momentum $\mathbf{p}.$ 

Assuming $p_x^2+p_y^2>0$, the inverse transformation is given
by the expressions
\begin{align}
\notag
     x&= \frac{2HT\cos\chi + L \sin\chi}{\sqrt{2MH}}, &    y&= \frac{2HT\sin\chi - L \cos\chi}{\sqrt{2MH}}, \\
    \label{new_coordinates1}
    p_x &= \sqrt{2MH}\cos\chi,  &   p_y &= \sqrt{2MH}\sin\chi.
\end{align}

Now, we would like to know what form the Moyal product (\ref{star-exp}) takes in the new coordinates (\ref{new_coordinates}). Recalculating the derivatives in formula \eqref{star-exp} is a difficult task.
A much more  efficient way to achieve this goal is to apply the Fedosov algorithm ~\cite{Fedosov1, Fedosov2}, which requires equipping our phase space $(\mathbb{R}^4,\,\omega)$ with a  symplectic connection $\gamma$ (see \cite{symplectic}). A parallel transport performed with this connection preserves the symplectic form $\omega$.  The symplectic manifold together with the symplectic connection  $(\mathbb{R}^4,\, \omega, \gamma)$ is known as the Fedosov manifold. 

The Fedosov construction is an advanced geometrical tool that can be applied even to curved manifolds \cite{gadella1}. However, on flat manifolds it is also helpful. Moreover, it is relatively easy to write a computer programme for this algorithm.

 We postulate that in the chart $(x,\,y,\,p_x,\,p_y)$ all the symplectic connection $\gamma$ coefficients vanish
\begin{align}
\label{gamma=0}
    \gamma^i_{\,\,\,jk} &= 0, &  i,\,j,\,k &= 1,\,2,\,3,\,4 .  
\end{align}
Under the change of coordinates 
\begin{equation*}
    (Q^a) := (x,\,y,\,p_x,\,p_y) \to (\widetilde{Q}^a):= (T, \,\chi,\,H,\,L),
\end{equation*}
 due to the condition (\ref{gamma=0}) the symplectic connection coefficients $\gamma_{ijk}$ transform according to the rule \cite{Gadella}
\begin{align}
\label{jt1}
    \widetilde{\gamma}_{ijk} &= \omega_{rd}\frac{\partial Q^r}{\partial \widetilde{Q}^i}\frac{\partial^2 Q^d}{\partial \widetilde{Q}^j \partial \widetilde{Q}^k}, & i,\,j,\,k &= 1,\,2,\,3,\,4.
\end{align}
In expression \eqref{jt1}, the Einstein summation convention is used. The coefficients of the symplectic connection $\widetilde{\gamma}_{ijk}$ are symmetric in all the indices. Thus, the only non-zero coefficients of the symplectic connection are 
\begin{align}
\label{gam}
\widetilde{\gamma}_{122} &=  -2H,  &
\widetilde{\gamma}_{133} &=  -\frac{1}{2H},  &
\widetilde{\gamma}_{222} &=  -2 \, {L},  &
\widetilde{\gamma}_{233} &=  \frac{L}{2H^2},  &
\widetilde{\gamma}_{234} &=  -\frac{1}{2H}.
\end{align}

\subsection{Star-eigenvalue equations for the Hamiltonian and angular momentum}
We apply the Fedosov algorithm (see  Appendix A) to construct the Fedosov star product `$\star$' for the symplectic connection (\ref{gam}). In this case, the star product `$\star$' is simply  the Moyal product `$\star_{(\text{M})}$' in the coordinates  (\ref{new_coordinates}). Then, using the alrea\text{d}y constructed $\star$-product in Eqs. (\ref{i1})--(\ref{i4}) for $R_1 = H$ and $R_2 = L$,
we find that the Wigner function $W_{Em}(T,\,\chi,\,H,\,L)$ of energy $E > 0$ and angular momentum $m\hbar$, where $m\in\mathbb{R}$, fulfils the system of equations
\setcounter{orange}{1}
\renewcommand{\theequation} {\arabic{section}.\arabic{equation}\theorange}
\begin{numcases}{}
\label{i1A}
  HW_{Em} -\frac{\hbar^2}{4}\bigg(H\frac{\partial^2 W_{Em}}{\partial L^2} + \frac{1}{4H}\frac{\partial^2 W_{Em}}{\partial T^2}  \bigg)  = EW_{Em}, \\
     \addtocounter{orange}{1}
\addtocounter{equation}{-1}
   \notag
L W_{Em} -\frac{\hbar^2}{4}\bigg( 
2\frac{\partial W_{Em}}{\partial L} +
L\frac{\partial^2 W_{Em}}{\partial L^2} + 2H\frac{\partial^2 W_{Em}}{\partial H \partial L} + \frac{1}{4H}\frac{\partial^2 W_{Em}}{\partial T \partial \chi} \\
 \label{i2A}
- \frac{L}{4H^2}\frac{\partial^2 W_{Em}}{\partial T^2}\bigg) = m\hbar W_{Em},
\end{numcases}
with the additional conditions 
  \addtocounter{orange}{1}
\addtocounter{equation}{-1}
\begin{numcases}{}
   \label{i3A}
\frac{\partial W_{Em}}{\partial T}  &= 0, \\
      \addtocounter{orange}{1}
\addtocounter{equation}{-1}
\label{i4A} 
\frac{\partial W_{Em}}{\partial \chi} &= 0.
\end{numcases}

The general solution of Eqs. (\ref{i3A})--(\ref{i4A}) is a function depending exclusively on the variables $H$ and $L$. Therefore, Eqs. (\ref{i1A})--(\ref{i2A}) turn into 
\setcounter{orange}{1}
\renewcommand{\theequation} {\arabic{section}.\arabic{equation}\theorange}
\begin{numcases}{}
\label{WmE1}
HW_{Em} - \frac{\hbar^2}{4}H\frac{\partial^2 W_{Em}}{\partial L^2} = EW_{Em}, \\
 \addtocounter{orange}{1}
\addtocounter{equation}{-1}
\label{WmE2}
L W_{Em} - \frac{\hbar^2}{2}\left(\frac{\partial W_{Em}}{\partial L} + H\frac{\partial^2 W_{Em}}{\partial L \partial H} + \frac{L}{2}\frac{\partial^2 W_{Em}}{\partial L^2}\right) = m\hbar W_{Em}.
\end{numcases}

\renewcommand{\theequation} {\arabic{section}.\arabic{equation}}

In the case when $H >E $,
one can show that the general solution of Eq. (\ref{WmE1}) is of the form
\begin{equation}
\label{w1I}
    W_{Em}(T,\,\chi,\,H,\,L) = A_1(H)\text{exp}\left(\frac{2 L}{\hbar}\sqrt{\frac{H - E}{H}}\right) + A_2(H)\text{exp}\left(-\frac{2 L}{\hbar}\sqrt{\frac{H - E}{H}}\right).
\end{equation}
The functions $A_1(H)$ and $A_2(H)$ are arbitrary but real. The absolute value of the solution (\ref{w1I}) increases rapidly for $L \rightarrow \pm\infty$, so it cannot be a Wigner function. Therefore, in the region
$H > E $ the only physically acceptable solution is 
\begin{equation}
    W_{Em}(T,\,\chi,\,H,\,L) = 0.
\end{equation}

In the case when $0< H < E$, the general real solution of Eq. (\ref{WmE1}) is of the form
\begin{equation}
\label{w2A}
      W_{Em}(T,\,\chi,\,H,\,L) = B_1(H)\cos\bigg(\frac{2 L}{\hbar}\sqrt{\frac{E - H}{H}}\bigg) + B_2(H)\sin\bigg(\frac{2 L}{\hbar}\sqrt{\frac{E - H}{H}}\bigg),
\end{equation}
where $B_1(H)$ and $B_2(H)$ are arbitrary real functions of $H$. Substituting the solution (\ref{w2A}) into Eq. (\ref{WmE2}), we obtain a differential equation for the functions $B_1(H)$ and $B_2(H)$. Solving it, we find that the complete solution of  (\ref{WmE1})--(\ref{WmE2}) for all $H>0$ is of the form 
\be
\label{jj1}
 W_{Em}(T,\,\chi,\,H,\,L) =  \frac{Y(E-H)N_{Em}}{\sqrt{H(E - H)}} \cos\left(\frac{2 L}{\hbar}\sqrt{\frac{E - H}{H}} - 2m\arccos\sqrt{\frac{H}{E}} + D_{Em}\right),
\ee
where $N_{Em}, \, D_{Em}$ are some real constants that may depend on $E$ and $m$. The symbol $Y(x)$ denotes the Heaviside step function
\begin{align}
    Y(x) := \begin{cases}
        1, & x >0, \\
        0, & x < 0.
    \end{cases}
\end{align}

It turns out that the function (\ref{jj1}) satisfies Eq. (\ref{WmE2}) in a distributional sense only if $D_{Em} = a \pi$, where $a \in \mathbb{Z}$. Let us put $D_{Em} = 0$. Thus, for all $E > 0 $ and $m \in \mathbb{R}$ we obtain 
\begin{align}
\label{WmE-final}
     W_{Em}(T,\,\chi,\,H,\,L) =  \frac{Y\left(E-H\right)N_{Em}}{\sqrt{H(E - H)}} \cos\left(\frac{2 L}{\hbar}\sqrt{\frac{E - H}{H}} - 2m\arccos\sqrt{\frac{H}{E}}\right).
\end{align}
The Wigner function $W_{Em}$  written in the form \eqref{WmE-final} was first derived in \cite{Schleich0, Schleich}. 

\subsection{Quantisation of the magnetic quantum number}
Let us prove that the  solution (\ref{WmE-final}) is a Wigner function only if $m \in \mathbb{Z}$. We will demonstrate this in two ways.


We first discuss a numerical method based on the probabilistic interpretation of quantum mechanics. Let us introduce the coordinates $(r,\,\phi,\,p,\,\chi)$, where  denote the polar coordinates of the position vector $\mathbf{r}$ and the polar coordinates of the momentum vector $ \mathbf{p}$, respectively. These new coordinates are not symplectic. The elementary volume
$\text{d}T \text{d}\chi \text{d}H \text{d}L$ transforms into $ r\text{d}r \text{d}\phi  p \text{d}p \text{d}\chi$. The energy is of the form $E = \frac{ p^2_0}{2M}$, where $p_0$ denotes the length of some momentum vector.   Then the function (\ref{WmE-final}) turns into
\begin{align}
    \label{Wmk0_1}
W_{E m}(r,\,\phi,\,p,\,\chi)  =\frac{Y(p_0 - p)  2M N_{Em}}{p\sqrt{p_0^2 - p^2}}     
     \cos\left(\frac{2r\sqrt{p_0^2 - p^2}}{\hbar}\sin(\chi - \phi) - 2m\arccos{\frac{p}{p_0}}\right).
\end{align}

For an arbitrary fixed  $m\in\mathbb{R}$, we define the integral as
\begin{align}
\label{Pmk0_kchi}
   P_{Em}(r,\,\phi) := \int^\infty_0 p\text{d}p  \int^{2\pi}_0 \text{d}\chi   W_{Em}(r,\,\phi,\,p,\,\chi).
\end{align}
If $W_{Em}$ is a Wigner function, the integral \eqref{Pmk0_kchi} represents the spatial density of probability. Substituting (\ref{Wmk0_1}) into (\ref{Pmk0_kchi}) and integrating over $\chi$, we obtain
\begin{align}
    P_{Em}(r,\,\phi) &= 4\pi MN_{Em}\int^{p_0}_0 \text{d}p \frac{1}{\sqrt{p^2_0 - p^2}} \cos\left(2m\arccos\frac{p}{p_0}\right)J_0\left(\frac{2r\sqrt{p^2_0 - p^2}}{\hbar}\right).
\end{align}
By $J_n(x), \; n \in \mathbb{Z}$, we mean the Bessel function of the first kind. One can see that the function $ P_{Em}$ does not depend on $\phi$.  We introduce a new variable $\theta$ fulfilling the conditions
\begin{align}
  p &= p_0\cos\theta,  & \theta &\in \left(0, \frac{\pi}{2}\right)
\end{align}
and we obtain
\begin{align}
\label{Pmk01}
    P_{Em}(r,\,\phi) &= 4\pi MN_{Em}\int^{\frac{\pi}{2}}_0 \text{d}\theta \cos(2m\theta)J_0 \left(\frac{2p_0r\sin\theta}{\hbar} \right).
\end{align} 
We are going to discuss two cases for the magnetic quantum number $m$: non-integer and integer.

Firstly, we examine the case when $m \in \mathbb{Z}$. In this case, the integral (\ref{Pmk01}) simplifies to
\begin{equation}
\label{PemZ}
    P_{Em}(r,\,\phi) = 2\pi^2 MN_{Em} J^2_{m}\left(\frac{p_0r}{\hbar}\right).
\end{equation}
Moreover, we can determine the normalisation constant $N_{Em}$ by postulating the normalisation condition (\ref{normalisation_def}) for $m,\,\widetilde{m}\in\mathbb{Z}$
\begin{equation}
\label{WEmWtEtm}
    (W_{Em} \star_{\text{(M)}}W_{\widetilde{E}\widetilde{m}})(T,\,\chi,\,H,\,L) =  \frac{1}{(2\pi \hbar)^2 M}\delta(E - \widetilde{E})\delta_{m\widetilde{m}}W_{Em}(T,\,\chi,\,H,\,L).
\end{equation}
Applying  the result \eqref{WEmm1B_final} from Appendix \ref{appendixB}, 
we obtain
\begin{equation}
\label{NEm1}
    N_{Em} = \frac{1}{4\pi^3M\hbar^2}.
\end{equation}
Hence, the solution written in the form (\ref{Wmk0_1}) with the normalisation constant (\ref{NEm1}) coincides with the result presented in \cite{Schleich}.

\begin{figure}[H]
\centering
\includegraphics[scale = 0.5]{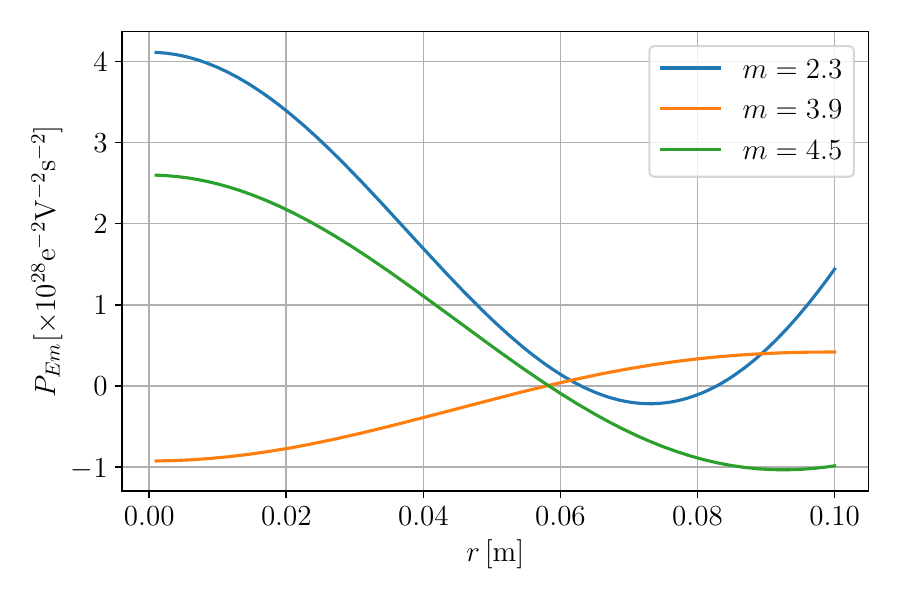}
 \caption{The numerical integral $P_{Em}(r,\,\phi)$, given by the formula (\ref{Pmk01}), as a function of radius $r$ for energy $E = 100\,\mathrm{eV}$, mass $M = 9\cdot 10^{-31}\,\mathrm{kg}$ and three  non-integer magnetic quantum numbers $m$. The normalisation constant $N_{Em}$ is chosen as (\ref{NEm1}).}
 \label{plot}
\end{figure}

Let us assume $m \in \mathbb{R}\setminus\mathbb{Z}$. Unfortunately, in this case, we do not know the closed form of the integral (\ref{Pmk01}). We choose the  value of the coefficient $N_{Em}$ as (\ref{NEm1}) and perform numerical calculations. In Fig. \ref{plot}, one can see  that the integral (\ref{Pmk01}) for a non-integer quantum magnetic number $m$ becomes negative for some radii $r$. This observation contradicts the interpretation of $ P_{Em}(r,\,\phi)$ as a spatial density of probability.

The second method is analytical, based on considerations presented in Appendix B. We prove that the normalisation condition (\ref{WEmm1B_final}) holds only for $m,\,\widetilde{m} \in \mathbb{Z}$ and the normalisation constant equals \eqref{NEm1}.

\begin{figure}[H]
\centering
\begin{subfigure}[t]{.5\textwidth}
  \includegraphics[width =\linewidth]{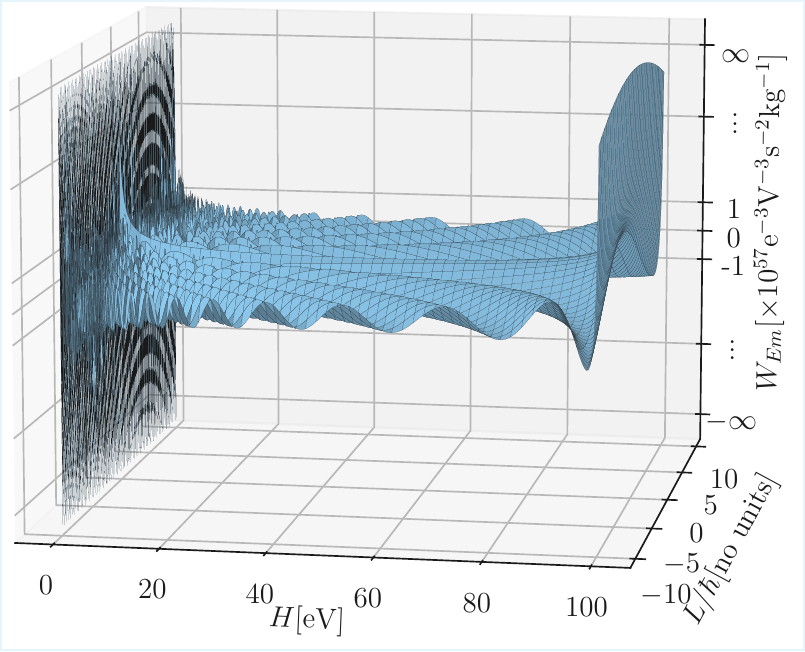}
  \caption{}
\end{subfigure}%
\begin{subfigure}[t]
{.5\textwidth}
\includegraphics[width=\linewidth]{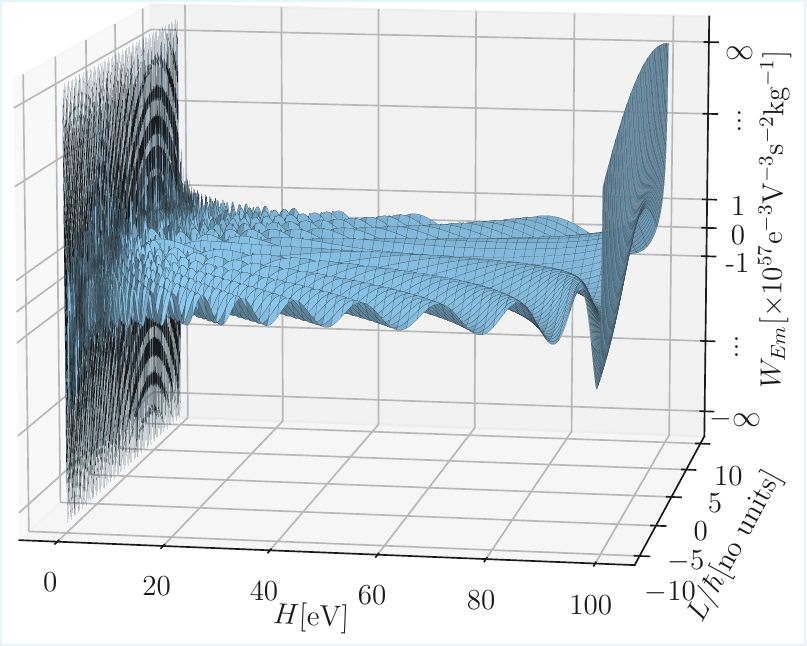}
  \caption{}
\end{subfigure}%
\caption{The Wigner function  $W_{Em}(T,\,\chi,\,H,\,L)$, given by the formula (\ref{WmE-final}) for energy $E = 100\,\mathrm{eV}$, mass $M = 9\cdot 10^{-31}\,\mathrm{kg}$ and  magnetic quantum number (a) $m=0$, (b) $m=5$. The normalisation constant $N_{Em}$ equals (\ref{NEm1}).}
\label{rys23}
\end{figure}

Figure \ref{rys23} shows two examples of Wigner functions $W_{Em}.$ As expected, they take both positive and negative values. For a fixed angular momentum $L \neq 0$ and variable $H \to 0^+$, the limit of the function $W_{Em}$ does not exist. However, if $H \to E^-$ then for arbitrary fixed $L$ the limit of $W_{Em}$ equals $ +\infty$.

\subsection{Semiclassical approximation}
Let us discuss the classical limit  of the Wigner function $W_{Em}$ given by the formula  (\ref{WmE-final}). Our considerations are analogous to those presented in \cite{semiclassical}.
 The classical limit is understood  as
\begin{align}
\label{WEmclass}
    W^{\text{class}}_{EJ}(T,\,\chi,\,H,\,L) := \frac{1}{(2 \pi \hbar)^2}\lim_{\substack{\hbar\to0\\m\to\infty\\ J = m\hbar}} (2 \pi \hbar)^2 W_{Em}(T,\,\chi,\,H,\,L),
\end{align}
where $J$ denotes the value of the angular momentum. The limits $\hbar \to 0 $ and $m \to +\infty$ in fact are not independent, since their product $m \hbar=J$ remains constant and much greater than $\hbar$. We understand these limits as indicators of tendencies rather than as limits in the strict mathematical sense.  

The reason why  we exclude the term $\frac{1}{(2 \pi \hbar)^2}$ from the limit $\hbar \to 0 $ is related to a fact known from statistical mechanics. Namely, to find the correct number of quantum states in the quasiclassical approximation one assumes that the volume of a single state equals  $(2 \pi \hbar)^2.$

 With the use of property \eqref{ident}, one can show that for an arbitrary function $F$ the  identity holds
\begin{align}
\label{identF}
    \frac{Y(E-H)}{\sqrt{H(E-H)}}F\bigg(\sqrt{\frac{E-H}{H}}\bigg) = \int^\infty_0 \frac{2 du}{1+u^2}\delta\bigg(H - \frac{E}{1+u^2}\bigg)F(u).
\end{align}
The Dirac delta $\delta\bigg(H - \frac{E}{1+u^2}\bigg)$ is a linear functional with respect to the variable $H.$ 

Thus, since the function is even, one observes that
\begin{align}
\label{Wclass1}
        W^{\text{class}}_{EJ}(T,\,\chi,\,H,\,L) =\frac{1}{(2 \pi \hbar)^2}  \lim_{\substack{\hbar\to0\\m\to\infty\\ J = m\hbar}}\frac{1}{\pi M }\int^\infty_{- \infty} \frac{ du}{1+u^2}\delta\bigg(H - \frac{E}{1+u^2}\bigg)\cos\bigg(\frac{2}{\hbar}[L u - J \arctan(u)]\bigg).
\end{align}
Since $J/\hbar=m$ and $m$ is large, the function $\cos\Big(2m \arctan(u)\Big)$ oscillates rapidly and its contribution to the integral is significant only for $u \approx 0.$
However, in that region the fraction $\frac{1}{1+u^2} \approx 1.$
Therefore, we may approximate the formula \eqref{Wclass1} by
\[
\frac{1}{4\pi^3M \hbar^2}\int^\infty_0  2du\delta\big(H - E\big)\cos\bigg(\frac{2}{\hbar}[u(L  - J)]\bigg)= \frac{1}{4\pi^3M \hbar^2}\int^\infty_{-\infty}  du\delta\bigg(H - E\bigg)\exp\bigg(\frac{2i}{\hbar}[u(L  - J)]\bigg).
\]
As expected,
\be
\label{xx1}
 W^{\text{class}}_{EJ}(T,\,\chi,\,H,\,L) = 
 \frac{1}{ (2 \pi\hbar)^2 M}
 \delta\big(H - E\big) \delta \left(\frac{L-J}{\hbar} \right)=
 \frac{\hbar}{ (2 \pi\hbar)^2 M}
 \delta\big(H - E\big) \delta \big(L-J \big).
 \ee
 Indeed, in the classical limit the Wigner function turns into a homogeneous  density of probability for a classical particle with energy $E$ and angular momentum $J.$ Thus,
 $W^{\text{class}}_{EJ}(T,\,\chi,\,H,\,L)$ is nonnegative---in contrast to the  original Wigner function $ W_{Em}(T,\,\chi,\,H,\,L).$  The Planck constant in the numerator is necessary in order to obtain the proper unit of the function $W^{\text{class}}_{EJ}(T,\,\chi,\,H,\,L).$

Formula \eqref{xx1} explains why the Wigner function \eqref{WmE-final} is not bounded for $H \to E^{-}.$ As we proved earlier, for $H>E$ the Wigner function $W_{Em}(T,\,\chi,\,H,\,L)$ vanishes. Then, in order to ensure that  in the classical limit the function
 $W_{Em}(T,\,\chi,\,H,\,L)$ transforms into 
the generalised function $ \delta\big(H - E\big) $, it must be that  $W_{Em}(T,\,\chi,\,H,\,L)$ tends to infinity for an arbitrary value of $L$ as $H$ approaches $E^{-}.$

In an analogous way, one should look at the oscillating dependence of the Wigner function \eqref{WmE-final} on the angular momentum. Indeed, treating the function $\cos(x)$ as a generalised function one obtains
\[
\lim_{s \to + \infty} \cos\left(s x \right)= 2 \pi \delta(x),
\]
which explains the origin of $\delta(L-J).$


\section{Cross-Wigner functions of eigenstates of energy and angular momentum}

\setcounter{equation}{0}

In the previous section, we derived the Wigner function $W_{Em}(\mathbf{r},\,\mathbf{p})$, given by Eq. (\ref{WmE-final}). This function represents the state $|\Psi_{Em}\rangle$ with a fixed energy $E >0$ and angular momentum $m\hbar$, where $m\in \mathbb{Z}$.  

We will denote the cross-Wigner function of the  states $|\Psi_{Em}\rangle$ and $|\Psi_{Em'}\rangle$ as $W_{Emm'}(\mathbf{r},\,\mathbf{p})$. It is defined for each $E>0$ and $m, m' \in \mathbb{R}$  as a solution to the following system of star-eigenvalue equations 
\setcounter{orange}{1}
\renewcommand{\theequation} {\arabic{section}.\arabic{equation}\theorange}
\begin{numcases}{}
\label{i1B}
   ( H \star_{(\text{M})}W_{E m m'}) (\mathbf{r},\,\mathbf{p}) = E W_{E m m'} (\mathbf{r},\,\mathbf{p}), \\ 
\addtocounter{orange}{1}
\addtocounter{equation}{-1}
\label{i2B}
   ( L \star_{(\text{M})}W_{E m m'})(\mathbf{r},\,\mathbf{p}) = m\hbar W_{E m m'}(\mathbf{r},\,\mathbf{p}), \\   \addtocounter{orange}{1}
\addtocounter{equation}{-1}
\label{i3B}
    (W_{E m m'}\star_{(\text{M})} L) (\mathbf{r},\,\mathbf{p}) = m' \hbar W_{E m m'}(\mathbf{r},\,\mathbf{p}),
\end{numcases}
fulfilling the commutation relation
\addtocounter{orange}{1}
\addtocounter{equation}{-1}
\begin{align}
\label{i4B}
\{H,\,W_{E m m'} \}_{(\text{M})}(\mathbf{r},\,\mathbf{p}) &= 0
\end{align}
and satisfying the generalised normalisation condition
\addtocounter{orange}{1}
\addtocounter{equation}{-1}
\begin{align}
\label{i5B}
\big( W_{Emm'}\star_{(\text{M})}W_{\widetilde{E}\widetilde{m}\widetilde{m}'}\big)  (T,\,\chi,\,H,\,L)&= \frac{1}{(2\pi\hbar)^2 M }\delta(E - \widetilde{E})\delta_{m'\widetilde{m}}W_{Em\widetilde{m}'}(T,\,\chi,\,H,\,L).
\end{align}
\renewcommand{\theequation} {\arabic{section}.\arabic{equation}}
Thus, we do not require the functions $W_{E m m'} (\mathbf{r},\,\mathbf{p})$ to be real.  From 
 Eqs. \eqref{i2B}--\eqref{i3B} and the property $\overline{A \star_{\text{(M)}} B} = \overline{B}\star_{\text{(M)}}\overline{A}$ it follows that
\be
\label{N1}
W_{E m m'} (\mathbf{r},\,\mathbf{p}) = \overline{W}_{E m' m} (\mathbf{r},\,\mathbf{p}).
\ee

In the coordinates $(T,\,\chi,\,H,\,L)$ defined by the relations (\ref{new_coordinates}), the formulas  (\ref{i1B}) and (\ref{i4B}) take the same form as Eqs. (\ref{i1A}) and (\ref{i3A}), respectively. Therefore, we postulate that
\begin{align}
\notag
    &W_{Emm'}(T,\,\chi,\,H,\,L) = \\ \label{w1IB} &Y(E-H) \bigg(B_1(H,\,\chi)\cos\bigg(\frac{2 L}{\hbar}\sqrt{\frac{E - H}{H}}\bigg) + B_2(H,\,\chi)\sin\bigg(\frac{2 L}{\hbar}\sqrt{\frac{E - H}{H}}\bigg) \bigg).
\end{align}
In order to find functions $B_1(H,\,\chi)$ and $B_2(H,\,\chi)$, we substitute the solution (\ref{w1IB}) into Eqs. (\ref{i2B}) and (\ref{i3B}). Then, using linear independence of $\sin\bigg(\frac{2 L}{\hbar}\sqrt{\frac{E - H}{H}}\bigg)$ and $\cos\bigg(\frac{2 L}{\hbar}\sqrt{\frac{E - H}{H}}\bigg) $, one can deduce that
\setcounter{orange}{1}
\renewcommand{\theequation} {\arabic{section}.\arabic{equation}\theorange}
\begin{numcases}{}
\label{i1C}
  (m - m')B_1 + i \frac{\partial B_1}{\partial \chi} = 0, \\ 
\addtocounter{orange}{1}
\addtocounter{equation}{-1}
\label{i2C}
     (m - m')B_2 + i \frac{\partial B_2}{\partial \chi} = 0, \\   \addtocounter{orange}{1}
\addtocounter{equation}{-1}
\label{i3C}
  \sqrt{(E - H)H}(m + m')B_1 + (E - 2H)B_2 + 2(E - H)H\frac{\partial B_2}{\partial H} = 0, \\
    \addtocounter{orange}{1}
\addtocounter{equation}{-1}
\label{i4C}
     -\sqrt{(E - H)H}(m + m')B_2 + (E - 2H)B_1 + 2(E - H)H\frac{\partial B_1}{\partial H} = 0.
\end{numcases}
\renewcommand{\theequation} {\arabic{section}.\arabic{equation}}
Finally,
\begin{align}
\notag
    W_{Emm'}(T,\,\chi,\,H,\,L) =  \frac{Y\left(E-H\right)N_{E m m'}}{\sqrt{H(E - H)}} e^{i(m - m')\chi} \\ 
    \label{WEmm1}
    \times \cos\left(\frac{2 L}{\hbar}\sqrt{\frac{E - H}{H}} - (m + m')\arccos\sqrt{\frac{H}{E}}\right).
\end{align}

 In Appendix B, we show that the normalisation condition holds only if $m,\,m',\,\widetilde{m},\,\widetilde{m}' \in \mathbb{Z}$ (compare with Eq. (\ref{norm1})),
\begin{align}
     \big( W_{Emm'}\star_{(\text{M})}W_{\widetilde{E}\widetilde{m}\widetilde{m}'}\big)(T,\,\chi,\,H,\,L)  
   \label{norm1A} = \frac{\pi N_{Emm'}N_{\widetilde{E}\widetilde{m}\widetilde{m}'}}{N_{Em\widetilde{m}'}} \delta(E - \widetilde{E}) W_{Em\widetilde{m}'}(T,\,\chi,\,H,\,L)\delta_{m'\widetilde{m}}.
\end{align}

Here, $N_{E m m'}$ is a normalisation constant that satisfies the relation \eqref{i5B} and also the condition
\begin{align}
\label{NEmm1=NEm1m}
    N_{E m m'} = \overline{N}_{E m' m},
\end{align}
due to Eq. (\ref{N1}).

Comparing (\ref{i5B}) and (\ref{norm1A}), we can deduce that  
\begin{align}
    N_{Emm'} &= \frac{1}{4\pi^3 M \hbar^2} z^{m-m'},
\end{align}
where $z \in \mathbb{C}\setminus\{0\}$ is such that Eq. (\ref{NEmm1=NEm1m}) holds---i.e. $z$ has to satisfy the requirement
\begin{equation}
\label{Cond1}
    z^{m - m'} = \overline{z^{m'-m}}.
\end{equation}
Using the property $|z^n| = |z|^n$ for $n\in\mathbb{Z}$, Eq. (\ref{Cond1}) implies that $|z| = 1$.
Therefore, the normalisation constant can be determined up to a phase factor $\alpha\in\mathbb{R}$ as
\begin{equation}
\label{NEmm1}
    N_{Emm'} = \frac{1}{4\pi^3M\hbar^2}\exp(i(m-m')\alpha).
\end{equation}

\section{Phase space representation of Jacobi-Anger expansion}
\setcounter{equation}{0}
In this section, we derive an expansion of the Wigner function for Cartesian momentum in terms of the cross-Wigner functions for energy and angular momentum. From a physical point of view, this is a problem of superposition in phase-space quantum mechanics. In Hilbert-space quantum mechanics, such an expansion is known as the Jacobi-Anger expansion. 

We want to find the coefficients $C_{\widetilde{m}\widetilde{m}'}$ such that
\begin{equation}
\label{expansion}
    W_{\widetilde{p}_{x0}\widetilde{p}_{y0}}(T,\,\chi,\,H,\,L) = \sum^\infty_{\widetilde{m} = -\infty}  \sum^\infty_{\widetilde{m}' = -\infty}  C_{\widetilde{m}\widetilde{m}'}W_{\widetilde{E}\widetilde{m}\widetilde{m}'}(T,\,\chi,\,H,\,L).
\end{equation}
The symbol $W_{\tilde{p}_{x0}\tilde{p}_{y0}}$ denotes the Wigner function with definite Cartesian momenta, given by
 \begin{equation}
 \label{Wpx0py0}
     W_{\tilde{p}_{x0}\tilde{p}_{y0}}(x,\,y,\,p_x,\,p_y) := \frac{1}{(2\pi\hbar)^2}\delta(p_x - \tilde{p}_{x0})\delta(p_y - \tilde{p}_{y0}),
 \end{equation}
 where $\tilde{p}_{x0},\,\tilde{p}_{y0}\in\mathbb{R}$ denote eigenvalues of the Cartesian components of the momentum. In the coordinates (\ref{new_coordinates}), this function takes the form
\begin{equation}
\label{Wpx0py0_TchiHL}
    W_{\tilde{p}_{x0}\tilde{p}_{y0}}(T,\,\chi,\,H,\,L) =  N_{\tilde{p}_{x0}\tilde{p}_{y0}}\delta(H - \tilde{E})\delta(\chi-\tilde{\chi}_0).
\end{equation}
The parameters in Eq. \eqref{Wpx0py0_TchiHL} are equal to
\begin{align}
\label{Ntildepxpy}
  N_{\tilde{p}_{x0}\tilde{p}_{y0}} &:= \frac{1}{(2\pi\hbar)^2 M}, &  \tilde{E} &:= \frac{\tilde{p}^2_{x0}+\tilde{p}^2_{y0}}{2M}, & \tilde{\chi}_0 &:= \text{atan2}(\tilde{p}_{y0},\,\tilde{p}_{x0}).
\end{align}

Expression (\ref{expansion}) represents the state $|\Psi_{\widetilde{p}_{x0}\widetilde{p}_{y0}}\rangle$ as a superposition of states $\{|\Psi_{\widetilde{E} m}\rangle\}_{m\in\mathbb{Z}}$ with $\widetilde{E} = \frac{\widetilde{p}^2_{x0}+\widetilde{p}^2_{y0}}{2M}$. We multiply both sides of Eq. (\ref{expansion}) by $W_{Emm'}$ in the sense of the Moyal product
\begin{equation}
\label{expansion1}
(W_{Emm'}\star_{(\text{M})}W_{\widetilde{p}_{x0}\widetilde{p}_{y0}})(T,\,\chi,\,H,\,L) = \underbrace{\sum^\infty_{\widetilde{m} = -\infty}  \sum^\infty_{\widetilde{m}' = -\infty}  C_{\widetilde{m}\widetilde{m}'}(W_{Emm'}\star_{(\text{M})}W_{\widetilde{E}\widetilde{m}\widetilde{m}'})(T,\,\chi,\,H,\,L)}_{=: Q}.
\end{equation}

In Appendix \ref{appendixC}, we prove that the LHS of Eq. (\ref{expansion1}) equals
\begin{align*}
    &\big( W_{Emm'}\star_{(\text{M})}W_{\widetilde{p}_{x0}\widetilde{p}_{y0}}\big)(T,\,\chi,\,H,\,L) = N_{Emm'}N_{\widetilde{p}_{x0}\widetilde{p}_{y0}}
    \frac{ Y\big(\cos(\chi - \widetilde{\chi}_0) \big)  }{\cos^2(\chi -\widetilde{\chi}_0 )}  e^{2im \chi}   e^{-i(m + m')\widetilde{\chi}_0 }
    \\ \tagg \label{WEmm1Wpx0py0_finalC1} &
     \times 
     \exp\bigg(-\frac{2i}{\hbar}L \tan(\chi -\widetilde{\chi}_0)\bigg) \delta \bigg(E-\widetilde{E}\bigg)
     \delta \bigg(E - \frac{H}{\cos^2(\chi - \widetilde{\chi}_0)} \bigg).
\end{align*} 

Let us focus on the RHS of Eq. (\ref{expansion1}), denoted as $Q$. We substitute the normalisation condition (\ref{i5B}), the normalisation constant (\ref{Ntildepxpy}) and the explicit form of $W_{Emm'}(\mathbf{r},\,\mathbf{p})$ given by Eq. (\ref{WEmm1}). We postulate that
\begin{equation}
C_{\widetilde{m}\widetilde{m}'} := \frac{1}{2\pi}\frac{N_{Em\widetilde{m}}}{N_{Em\widetilde{m}'}}e^{i(-\widetilde{m} + \widetilde{m}')\widetilde{\chi}_0} = \frac{1}{2\pi} e^{i(-\widetilde{m}+\widetilde{m}')\alpha}e^{i(-\widetilde{m} + \widetilde{m}')\widetilde{\chi}_0}.
\end{equation} 
Then, 
\begin{align*}
    &Q = N_{Emm'}  N_{\widetilde{p}_{x0}\widetilde{p}_{y0}}\delta(E - \widetilde{E}) \sum^\infty_{\widetilde{m}' = -\infty}  \frac{1}{2\pi}e^{i(-m' + \widetilde{m}')\widetilde{\chi}_0}  
\\ \tagg &\times \frac{Y(E - H)}{\sqrt{H(E - H)}}  e^{i(m - \widetilde{m}')\chi}\cos\bigg(\frac{2 L}{\hbar}\sqrt{\frac{E - H}{H}} - (m + \widetilde{m}')\arccos\sqrt{\frac{H}{E}}\bigg).
\end{align*}
Using the formula $\cos(x) = \frac{e^{ix} + e^{-ix}}{2}$, one can write $Q$ as
\begin{align*}
    &Q=N_{Emm'}  N_{\widetilde{p}_{x0}\widetilde{p}_{y0}}\delta(E - \widetilde{E}) \frac{1}{4\pi}\frac{Y(E - H)}{\sqrt{H(E - H)}}  e^{i(m\chi-m'\widetilde{\chi}_0)} \\ &\times \bigg\{
    \exp\bigg( i\frac{2 L}{\hbar}\sqrt{\frac{E - H}{H}}\bigg)\exp\bigg(-i m \arccos\sqrt{\frac{H}{E}}\bigg)  \sum^\infty_{\widetilde{m}' = -\infty}  
\exp\bigg( i\widetilde{m}'\bigg[-\chi+\widetilde{\chi}_0 -\arccos\sqrt{\frac{H}{E}}\bigg]\bigg) \\ &+ \exp\bigg(-i\frac{2 L}{\hbar}\sqrt{\frac{E - H}{H}}\bigg)\exp\bigg(i m \arccos\sqrt{\frac{H}{E}}\bigg)  \sum^\infty_{\widetilde{m}' = -\infty}  
\exp\bigg( i\widetilde{m}'\bigg[-\chi+\widetilde{\chi}_0 +\arccos\sqrt{\frac{H}{E}}\bigg]\bigg) 
    \bigg\}.
\end{align*}
Using the identity $\sum^\infty_{n=-\infty}e^{inx} = 2\pi\sum^\infty_{k = -\infty}\delta(x + 2k\pi)$ for $x\in\mathbb{R}$, we obtain
\begin{align*}
    &Q=N_{Emm'}  N_{\widetilde{p}_{x0}\widetilde{p}_{y0}}\delta(E - \widetilde{E}) \frac{1}{2}\frac{Y(E - H)}{\sqrt{H(E - H)}}  e^{i(m\chi-m'\widetilde{\chi}_0)} \\ &\times \bigg\{
    \exp\bigg( i\frac{2 L}{\hbar}\sqrt{\frac{E - H}{H}}\bigg)\exp\bigg(-i m \arccos\sqrt{\frac{H}{E}}\bigg) \sum^\infty_
          {k=-\infty}\delta\bigg(-\chi + \widetilde{\chi}_0 -  \arccos\sqrt{\frac{H}{E}} + 2k\pi \bigg) \\ &+ \exp\bigg(-i\frac{2 L}{\hbar}\sqrt{\frac{E - H}{H}}\bigg)\exp\bigg(i m \arccos\sqrt{\frac{H}{E}}\bigg) \sum^\infty_
          {k=-\infty}\delta\bigg(-\chi + \widetilde{\chi}_0 +  \arccos\sqrt{\frac{H}{E}} + 2k\pi \bigg) 
    \bigg\}.
\end{align*}
The Dirac deltas lead to the following set of constraints:
\begin{equation}
\label{chimEq}
   - \chi + \widetilde{\chi}_0 \mp \arccos\sqrt{\frac{H}{E}} + 2k\pi = 0.
\end{equation}
The solution of Eq. (\ref{chimEq}) exists only if
\begin{enumerate}
\item[(i)] $\chi-\widetilde{\chi}_0 \in (3\pi/2,\,2\pi) \text{mod}(2\pi)$ for the `$-$' choice,
\item[(ii)] $\chi-\widetilde{\chi}_0 \in (0,\,\pi/2) \text{mod}(2\pi)$ for the `$+$' choice. 
\end{enumerate}
In both cases, the solution of Eq. (\ref{chimEq}) is of the form
\begin{equation}
    E = \frac{H}{\cos^2\big(\chi - \widetilde{\chi}_0\big)}.
\end{equation}
Then, using the coordinate-transformation rule for the Dirac delta with respect to $E$, we obtain
\begin{align*}
        \sum^\infty_
          {k=-\infty}\delta\bigg(&-\chi + \widetilde{\chi}_0 \mp  \arccos\sqrt{\frac{H}{E}} + 2k\pi \bigg) \\ \tagg &= 2 E \sqrt{\frac{E-H}{H}}\delta\bigg(E - \frac{H}{\cos^2\big(\chi - \widetilde{\chi}_0\big)}\bigg)Y\big(\cos\big(\chi-\widetilde{\chi}_0\big)\big)Y(\mp \sin(\chi-\widetilde{\chi}_0)).
\end{align*}
Therefore,
\begin{align*}
    &Q=N_{Emm'}  N_{\widetilde{p}_{x0}\widetilde{p}_{y0}}\delta(E - \widetilde{E}) \frac{E}{H}Y(E - H)e^{i(m\chi-m'\widetilde{\chi}_0)}Y\big(\cos\big(\chi-\widetilde{\chi}_0\big)\big) \delta\bigg(E - \frac{H}{\cos^2\big(\chi - \widetilde{\chi}_0\big)}\bigg) \\ &\times \bigg\{
    \exp\bigg( i\frac{2 L}{\hbar}\sqrt{\frac{E - H}{H}}\bigg)\exp\bigg(-i m \arccos\sqrt{\frac{H}{E}}\bigg)   Y(- \sin(\chi-\widetilde{\chi}_0)) \\ \tagg &+ \exp\bigg(-i\frac{2 L}{\hbar}\sqrt{\frac{E - H}{H}}\bigg)\exp\bigg(i m \arccos\sqrt{\frac{H}{E}}\bigg)  Y(+ \sin(\chi-\widetilde{\chi}_0))  
    \bigg\}.
\end{align*}
Finally, we can use the identity $f(x)\delta(x-a) = f(a)\delta(x-a)$ to recover Eq. (\ref{WEmm1Wpx0py0_finalC1}). Thus, we obtain
\begin{equation}
\label{final_expans}    W_{\widetilde{p}_{x0}\widetilde{p}_{y0}}(T,\,\chi,\,H,\,L) = \sum^\infty_{\widetilde{m} = -\infty}  \sum^\infty_{\widetilde{m}' = -\infty}  \frac{1}{2\pi}e^{i(-\widetilde{m}+\widetilde{m}')\alpha} e^{i(-\widetilde{m} + \widetilde{m}')\widetilde{\chi}_0}W_{\widetilde{E}\widetilde{m}\widetilde{m}'}(T,\,\chi,\,H,\,L).
\end{equation}

 In Hilbert-space quantum mechanics, the Jacobi-Anger expansion \cite{polar} holds for the wavefunctions $\psi_{Em}(r,\,\phi) = \frac{1}{\sqrt{2\pi}\hbar}J_m(k_0 r)e^{im\phi}$ (compare with \cite{Schleich, Varro}) and $\psi_{p_{x0}p_{y0}}(r,\,\phi) = \frac{1}{2\pi\hbar}e^{irk_0 \cos(\chi_0-\phi)}$  with $E = \frac{(\hbar k_0)^2}{2M}$ 
\begin{equation}
\label{JA_exp}
\psi_{p_{x0}p_{y0}}(r,\,\phi)  = \sum^\infty_{m = -\infty}\frac{1}{\sqrt{2\pi}}i^me^{-im\chi_0} \psi_{Em}(r,\,\phi).
\end{equation}

Thus, for $\alpha = -\pi/2$, the expansion (\ref{final_expans}) is compatible with Eq. (\ref{JA_exp}).


\section{Conclusions}
All considerations in the present paper are confined to the formalism of phase-space quantum mechanics.
We  derive the energy-angular-momentum Wigner function $W_{Em}$ and the cross-Wigner function $W_{Emm'}$ by solving the $\star_{\text{(M)}}$-eigenvalue equations. We show that only solutions with a quantised magnetic quantum number that satisfy the normalisation condition are physically acceptable.

It is worth highlighting that we  construct the Moyal $\star_{(\text{M})}$-product using the Fedosov algorithm adapted to the symplectic coordinates $(T,\,\chi,\,H,\,L)$. The Fedosov method is a powerful tool for dealing with non-Cartesian coordinates. We  prove that the coordinates $(T,\,\chi,\,H,\,L)$ are the most suitable for energy-angular-momentum $\star_{(\text{M})}$-eigenvalue problem because the equations $\{H,\,W_{Em}\}_{(\text{M})}=0$ and $\{L,\,W_{Em}\}_{\text{(M)}}=0$ are satisfied by  functions of the form $W_{Em}(T,\,\chi,\,H,\,L)=W_{Em}(H,\,L)$. This results from the fact that the Moyal bracket in this case reduces to the Poisson bracket.

The semiclassical approximation of the Wigner function $W_{Em}$ is also derived. It reveals the physical role of oscillatory behaviour in angular momentum, since it enables the Wigner function to collapse to the  Dirac delta with respect to the angular momentum $L.$

Finally, inspired by the Jacobi-Anger expansion in Hilbert-space quantum mechanics, we expand the Wigner function of Cartesian momentum $W_{\widetilde{p}_{x0}\widetilde{p}_{y0}}$ in the cross-Wigner functions of energy-angular momentum $W_{Emm'}$. This is achieved by calculating the products $W_{Emm'}\star_{\text{(M)}}W_{E\widetilde{m}\widetilde{m}'}$ and $W_{Emm'}\star_{\text{(M)}} W_{\widetilde{p}_{x0}\widetilde{p}_{y0}}$. An open question remains whether the proposed algorithm can be applied to other expansions known in the Hilbert-space formalism---e.g. to interbasis expansions of the 2D hydrogen atom \cite{hydrogen} or the Rayleigh expansion.

\section*{Acknowledgement}
This article was completed while the first author was a Doctoral Candidate at the Interdisciplinary Doctoral School at Lodz University of Technology, Poland. 

\section*{Conflict of interest statement} On behalf of all authors, the corresponding author states that there is no conflict of interest.

\section*{Data availability statement}
The datasets used and/or analysed during the current study are available from the corresponding author on reasonable request.

\section*{Author contribution statement}
All authors wrote the main manuscript text. Hubert Jóźwiak prepared figures. All authors reviewed the manuscript.

\section*{Funding declaration}
This research received no specific grant from any funding agency in the public, commercial, or not-for-profit sectors.

\appendix
\renewcommand{\thesection}{\Alph{section}}
\renewcommand{\theequation} {\Alph{section}.\arabic{equation}}
\setcounter{section}{0}
\setcounter{equation}{0}
\section{An adaptation of the construction of the Fedosov star product to flat symplectic manifolds}
This Appendix contains a  sketch of the algorithm \cite{Fedosov1, Fedosov2} for constructing a star product on an arbitrary Fedosov manifold. Some parts of the original Fedosov proposal are modified, since we deal exclusively with flat symplectic manifolds. As before, the Einstein summation convention is applied.

Let $(\mathcal{M},\,\omega, \gamma)$ be a $2n$-dimensional flat Fedosov manifold \cite{symplectic}---i.e. a $2n$-D manifold $\mathcal{M}$ equipped with a symplectic structure   $\omega$ and  a flat  connection $\gamma$  which is torsion free  and preserves  the symplectic form $\omega$. 

The Fedosov algorithm works only in the Darboux coordinates $(x^1, \ldots, x^{2n})$, but every symplectic manifold can be covered with a Darboux atlas. In these coordinates, the matrix $\omega_{ij}$ of the symplectic form and its inverse $\omega^{ij}$ are, respectively, equal to
\begin{align}
\label{xx}
    (\omega_{ij}) &= \begin{bmatrix}
        \mathbb{O} & \mathbb{I} \\
        -\mathbb{I} & \mathbb{O} 
    \end{bmatrix}, & (\omega^{ij}) &=  \begin{bmatrix}
        \mathbb{O} & -\mathbb{I} \\
        \mathbb{I} & \mathbb{O} 
    \end{bmatrix},
\end{align}
where the symbol $\mathbb{O}$ denotes the zero square matrix of dimension $n \times n$ and the symbol $ \mathbb{I}$ represents the 
$n \times n$ identity matrix.

The symplectic  connection $\gamma$ is determined by the coefficients $\gamma^i_{\,\,\,jk}$ or equivalently by $\gamma_{ijk}:=\omega_{ir}\gamma^r_{\,\,\,jk}$. In every Darboux chart $(x^1, \ldots, x^{2n})$  the coefficients $\gamma_{ijk}$ are symmetric in all indices
\begin{align}
    \gamma_{ijk} &= \gamma_{ikj} = \gamma_{jik} = \gamma_{jki} = \gamma_{kij} = \gamma_{kji}, &  i,\,j,\,k &= 1,\,\dots,\,2n.
\end{align}

At each point $x \in \mathcal{M}$ we introduce the Weyl algebra $(W_x,\,\circ,\,\mathbb{C}),$ whose elements are formal series in $\hbar$
\begin{equation}
    W_x := \left\{a=\sum^\infty_{k,\,l = 0} \hbar^k a_{k,i_1\dots i_l} y^{i_1}\dots y^{i_l}: (y^1,\,\dots,\,y^{2n})\in T_x\mathcal{M},\,  a_{k,i_1\dots i_l} \in (T^*_x\mathcal{M})^{\otimes_s l} \right\}
\end{equation}
and for which the  $\circ$-product equals
\begin{equation}
\label{circ}
    a \circ b := \sum^\infty_{k = 0}\left(\frac{i\hbar}{2}\right)^k\frac{1}{k!}\omega^{i_1j_1}\dots\omega^{i_kj_k}\frac{\partial^k a}{\partial y^{i_1}\dots\partial y^{i_k}}\frac{\partial^k b}{\partial y^{j_1}\dots\partial y^{j_k}}.
\end{equation}
We take a union of the algebras $(W_x,\,\circ,\,\mathbb{C})$ for all $x \in \mathcal{M}$ to obtain a bundle of formal Weyl algebras $(W,\,\circ,\,\mathbb{C})$ where
\begin{equation}
    W := \bigcup_{x\in\mathcal{M}} W_x
\end{equation}
with the fibrewise multiplication (\ref{circ}). 

In the Fedosov algorithm, the crucial role is played by differential forms on $\mathcal{M}$ with values in the bundle $W$---i.e. the following algebra $(\oplus^{2n}_{q=0}(W\otimes\Lambda^q) ,\,\bullet,\,\mathbb{C})$ where
\begin{align}
\notag
   W\otimes\Lambda^q   := \Bigg\{ \sum^\infty_{k,\,l = 0} \hbar^k a_{k,\,i_1\dots i_l, j_1\dots j_q}(x^1, \ldots, x^{2n})y^{i_1}\dots y^{i_{l}}\text{d}x^{j_1}\wedge\ldots\wedge\text{d}x^{j_q}:  \forall_{x\in\mathcal{M}}\\
    a_{k,\,i_l\dots i_l, j_1\dots j_q}(x^1, \ldots, x^{2n}) \in (T^*_x\mathcal{M})^{\otimes_s l}\otimes (T^*_x\mathcal{M})^{\otimes_a q},\, (y^1,\,\dots,\,y^{2n})\in T_x\mathcal{M}
    \Bigg\}.
\end{align}
From now on, we will omit variables $(x^1, \ldots, x^{2n})$ on the symplectic manifold $\mathcal{M}.$

The multiplication `$\bullet$' is defined by means of the exterior product of differentials $\text{d}x^i$ and the Weyl product (\ref{circ})
\begin{align}
\notag
 a \bullet b :=  \Bigg(&\sum^\infty_{k,\,l_1 = 0} \hbar^k a_{k,\,i_1\dots i_{l_1}, j_1\dots j_{q_1}} y^{i_1}\dots y^{i_{l_1}}\Bigg)\circ\Bigg(\sum^\infty_{k,\,l_2 = 0} \hbar^k b_{k,\,I_1\dots I_{l_2}, J_1\dots J_{q_2}} y^{I_1}\dots y^{I_{l_2}}\Bigg) \\
 &\times \text{d}x^{j_1}\wedge\ldots\wedge\text{d}x^{j_{q_1}} \wedge \text{d}x^{J_1}\wedge\ldots\wedge\text{d}x^{J_{q_2}}.
\end{align}
The commutator $[\cdot,\,\cdot]$ for any $a\in W\otimes \Lambda^{q_1}$ and $b\in W\otimes\Lambda^{q_2}$ is given by the formula
\begin{equation}
    [a,\,b] := a \bullet b- (-1)^{q_1q_2} b\bullet a.
\end{equation}

The  projection $P:W\otimes\Lambda^q\to Z\otimes\Lambda^q$ assigns to any $a \in W\otimes \Lambda^q$ its part not containing $y^i$'s
\begin{align}
\label{P} 
P(a) :=  \sum^\infty_{k = 0} \hbar^k a_{k, j_1\dots j_q}\text{d}x^{j_1}\wedge\ldots\wedge\text{d}x^{j_q}.
\end{align}
By $Z$ we mean the linear space of formal series over the ring of smooth functions $C^{\infty}({\cal M})$ in $\hbar$.

The symplectic  connection $\partial :W\otimes \Lambda^q \to W\otimes \Lambda^{q+1}$ acts   as
 \begin{equation}
 \label{D-def}
     \partial a := \text{d}a + \frac{i}{\hbar}[\Gamma,\,a],
 \end{equation}
 where $\text{d} := \text{d}x^i\wedge \frac{\partial}{\partial x^i}$ and
    $\Gamma := \frac{1}{2}\gamma_{ijk}y^iy^j\text{d}x^k$.

In the next step, we introduce an Abelian connection $\text{D}:W\otimes \Lambda^q \to W\otimes \Lambda^{q+1}$ related to the symplectic connection $\gamma_{ijk}$ and  given by the formula
\be
\text{D}a:= \partial a + \frac{i}{\hbar}[\omega_{ij}y^i \text{d}x^j,\,a].
\ee

The subalgebra $W_{\text{D}} \subset W$ of the Weyl algebra corresponding to the Abelian  connection $D$ is defined as a collection of flat sections
\begin{equation}
    W_{\text{D}} := \{a\in W: \text{D} a = 0 \}.
\end{equation}
One can show that the element $a \in W_\text{D}$ assigned to $P(a) \in Z$ is given uniquely as a solution of the recurrence equation 
\begin{equation}
\label{j1}
    a = P(a) + \delta^{-1}\partial a.
\end{equation}
We stress that the formula \eqref{j1} is valid only for  flat symplectic connections.

The symbol  $\delta^{-1}$ denotes a linear operator such that for each monomial \newline $y^{i_1} \dots y^{i_l}\text{d}x^{j_1}\wedge\ldots \wedge \text{d}x^{j_q}$ we obtain
\begin{align}
\notag
    &\delta^{-1}(y^{i_1}\dots y^{i_l}\text{d}x^{j_1}\wedge\ldots \wedge \text{d}x^{j_q}) := \\ &\begin{cases} \frac{1}{l 
+ q}\sum^q_{r = 1}(-1)^{r + 1}y^{j_r}y^{i_1}\dots y^{i_l}\text{d}x^{j_1}\wedge\ldots  \overset{\vee}{\text{d}x^{j_r}}\ldots\wedge \text{d}x^{j_q}&,\,l + q > 0, \\
0&,\,l+q = 0.
\end{cases}
\end{align}

One can notice that the map $\sigma: W_{\text{D}} \to Z$ defined for any $a\in W_{\text{D}}$ as
\begin{equation}
    \sigma(a) := P(a)
\end{equation}
is bijective. Thus, the inverse map $\sigma^{-1}$ exists and for any $f \in Z$  is determined by the recurrence equation 
\begin{equation}
    \sigma^{-1}(f) = f  + \delta^{-1}\partial \sigma^{-1}(f).
\end{equation}
Therefore, the Fedosov star product $\star:Z\times Z \to Z$ is constructed for each $f,\,g\in Z$ as 
\begin{equation}
\label{Fstar}
    f\star g := \sigma(\sigma^{-1}(f)\circ\sigma^{-1}(g))
\end{equation}
and on the manifold ${\cal M}= {\mathbb R}^{2n} $ is equal to the Moyal product $f \star_{(\text{M})} g $ of functions $f$ and $g.$

\renewcommand{\thesection}{\Alph{section}}
\renewcommand{\theequation} {\Alph{section}.\arabic{equation}}
\setcounter{section}{1}
\setcounter{equation}{0}
\section{Calculation of the Moyal product of functions \\ $(W_{Emm'}\star_{(\text{M})}W_{\widetilde{E}\widetilde{m}\widetilde{m}'})(T,\,\chi,\,H,\,L)$}
\label{appendixB}
The Moyal product of cross-Wigner functions is of the form
\begin{align*}
    &\big( W_{Emm'}\star_{(\text{M})}W_{\widetilde{E}\widetilde{m}\widetilde{m}'}\big)(T,\,\chi,\,H,\,L) = \frac{N_{Emm'}N_{\widetilde{E}\widetilde{m}\widetilde{m}'}}{(\pi \hbar)^4} \int^\infty_{-\infty}\text{d}T'\int^{ 2\pi}_{0}\text{d}\chi'\int^\infty_0 \text{d}H' \int^\infty_{-\infty}\text{d}L' \\ &\times \int^\infty_{-\infty}\text{d}T''\int^{ 2\pi}_{0}\text{d}\chi''\int^\infty_0 \text{d}H'' \int^\infty_{-\infty}\text{d}L''\frac{Y(E - H')}{\sqrt{H'(E - H')}} \frac{Y(\widetilde{E} - H'')}{\sqrt{H''(\widetilde{E} - H'')}}  e^{i(m - m')\chi'} e^{i(\widetilde{m} - \widetilde{m}')\chi''}
\\&\times\cos\left(\frac{2 L'}{\hbar}\sqrt{\frac{E - H'}{H'}} - (m + m')\arccos\sqrt{\frac{H'}{E}}\right) \\ &\times
 \cos\left(\frac{2 L''}{\hbar}\sqrt{\frac{\widetilde{E} - H''}{H''}} - (\widetilde{m} + \widetilde{m}')\arccos\sqrt{\frac{H''}{\widetilde{E}}}\right) \\
 &\times \exp\bigg(\frac{2i}{\hbar}\bigg\{ 
 2T\bigg[\sqrt{H H'}\cos(\chi - \chi') -\sqrt{HH''}\cos(\chi - \chi'')  \bigg] \\ &+ 2T' \bigg[\sqrt{H'H''}\cos(\chi' - \chi'') - \sqrt{HH'}\cos(\chi - \chi')\bigg]  \\ &+ 2T''\bigg[\sqrt{HH''}\cos(\chi - \chi'') - \sqrt{H'H''}\cos(\chi'-\chi'')\bigg] \\ &+ 
  L \bigg[\sqrt{\frac{H'}{H}}\sin(\chi - \chi') - \sqrt{\frac{H''}{H}}\sin(\chi - \chi'')\bigg] + L'\bigg[\sqrt{\frac{H}{H'}}\sin(\chi - \chi') + \sqrt{\frac{H''}{H'}}\sin(\chi' - \chi'')\bigg] \\ \tagg  &+ L''\bigg[\sqrt{\frac{H'}{H''}}\sin(\chi'-\chi'') - \sqrt{\frac{H}{H''}}\sin(\chi - \chi'')\bigg] \bigg\}\bigg).
\end{align*} 

First, we make the linear change of variables $\chi'\to\chi'-\chi$ and $\chi''\to\chi''-\chi$. Next, we integrate over $T'$ and $T''$, applying the identities $\int^\infty_{-\infty}\text{d}x e^{ikx} = 2\pi\delta(k)$, $\delta(ax) = \delta(x)/|a|$ and $f(x)\delta(x - b)=f(b)\delta(x - b)$, where $a\in\mathbb{R}\setminus\{0\}$ and $b\in\mathbb{R}$. Thus, we obtain

\begin{align*}
  &\big( W_{Emm'}\star_{(\text{M})}W_{\widetilde{E}\widetilde{m}\widetilde{m}'}\big)(T,\,\chi,\,H,\,L) = \frac{N_{Emm'}N_{\widetilde{E}\widetilde{m}\widetilde{m}'}}{4(\pi \hbar)^2} e^{i(m-m'+\widetilde{m} - \widetilde{m}')\chi}\int^{-\chi + 2\pi}_{-\chi}\text{d}\chi'\int^\infty_0 \text{d}H' \\ &\times  \int^\infty_{-\infty}\text{d}L' \int^{-\chi + 2\pi}_{-\chi}\text{d}\chi''\int^\infty_0 \text{d}H'' \int^\infty_{-\infty}\text{d}L''\frac{Y(E - H')}{\sqrt{H'(E - H')}} \frac{Y(\widetilde{E} - H'')}{\sqrt{H''(\widetilde{E} - H'')}}   e^{i(m - m')\chi' } e^{i(\widetilde{m} - \widetilde{m}')\chi''}\\ &\times
\cos\left(\frac{2 L'}{\hbar}\sqrt{\frac{E - H'}{H'}} - (m + m')\arccos\sqrt{\frac{H'}{E}}\right) \\&\times
 \cos\left(\frac{2 L''}{\hbar}\sqrt{\frac{\widetilde{E} - H''}{H''}} - (\widetilde{m} + \widetilde{m}')\arccos\sqrt{\frac{H''}{\widetilde{E}}}\right) \\
 &\times \exp\bigg(\frac{2i}{\hbar}\bigg\{  L \bigg[-\sqrt{\frac{H'}{H}}\sin(\chi') + \sqrt{\frac{H''}{H}}\sin( \chi'')\bigg] \\
 &+ L'\bigg[-\sqrt{\frac{H}{H'}}\sin(\chi') + \sqrt{\frac{H''}{H'}}\sin(\chi' - \chi'')\bigg] + L''\bigg[\sqrt{\frac{H'}{H''}}\sin(\chi'-\chi'') + \sqrt{\frac{H}{H''}}\sin(  \chi'')\bigg] \bigg\}\bigg)  \\ \tagg 
 \label{B2}
 &\times \delta\bigg(\sqrt{H' H''}\cos(\chi' - \chi'') - \sqrt{HH'}\cos(\chi')\bigg)\delta\bigg(\sqrt{H' H''}\cos(\chi' - \chi'') - \sqrt{HH''}\cos(\chi'')\bigg).
\end{align*}
Let us recall the identity \cite{Schwartz}
\begin{align}
\notag
    \iint_D \text{d}x_1 \dots  \text{d}x_n \delta(f_1(x_1,\,\dots,\,x_n))\dots \delta(f_n(x_1,\,\dots,\,x_n)) h(x_1,\,\dots,\,x_n) \\
    \label{ident}
    =  \sum_{(x_{10},\,\dots,\,x_{n0}) \in D_0}\bigg|\frac{\partial (f_1,\,\dots,\,f_n)}{\partial (x_1,\,\dots,\,x_n)}\bigg|^{-1}_{(x_{10},\,\dots,\,x_{n0})}h(x_{10},\,\dots,\,x_{n0}),
\end{align}
where 
    $D_0 := \{(x_1,\,\dots,\,x_n)\in D :f_1(x_1,\,\dots,\,x_n) = 0 \land \dots \land f_n(x_1,\,\dots,\,x_n) = 0\}$.
    
As $H, H'$ and $H''$ are positive, all of $\cos(\chi'- \chi''),\cos(\chi'), \cos(\chi'') $ must be different from $0.$
Therefore, the Dirac delta constraints in \eqref{B2} are satisfied by
\begin{align}
    H' &= H\frac{\cos^2(\chi'')}{\cos^2(\chi'-\chi'')}& &\land &
     H'' &= H\frac{\cos^2(\chi')}{\cos^2(\chi'-\chi'')},
\end{align}
under  two extra conditions:
\be
\label{dod1}
Y\bigg(\frac{\cos(\chi'')}{\cos(\chi'-\chi'')}\bigg) = Y\bigg(\frac{\cos(\chi')}{\cos(\chi'-\chi'')}\bigg)=1.
\ee
Observe that \eqref{dod1} implies the inequality
\[
\cos(\chi'-\chi'') \geq \sin(\chi') \sin(\chi'').
\]
One can also notice that neither $\chi'$ nor $\chi''$ equals $\pm \pi.$

  We integrate over $H'$ and $H''$, using the formula (\ref{ident}). Hence
\begin{align*}
     &\big( W_{Emm'}\star_{(\text{M})}W_{\widetilde{E}\widetilde{m}\widetilde{m}'}\big)(T,\,\chi,\,H,\,L) = \frac{N_{Emm'}N_{\widetilde{E}\widetilde{m}\widetilde{m}'}}{(\pi \hbar)^2}e^{i(m-m'+\widetilde{m}-\widetilde{m}')\chi} 
\\   
   &\times \int^{-\chi + 2\pi}_{-\chi}\text{d}\chi'\int^\infty_{-\infty}\text{d}L' \int^{-\chi + 2\pi}_{-\chi}\text{d}\chi''\int^\infty_{-\infty}\text{d}L''\frac{1}{\cos^{2}(\chi'-\chi'')} e^{i(m - m')\chi'} e^{i(\widetilde{m} - \widetilde{m}')\chi''}
\\
&\frac{\times Y\left(E-H\frac{\cos^{2}( \chi'')}{\cos^{2}(\chi'-\chi'')}\right)\, Y\left(\widetilde E-H\frac{\cos^{2}( \chi')}{\cos^{2}(\chi'-\chi'')}\right)Y\bigg(\frac{\cos(\chi')}{\cos(\chi'-\chi'')}\bigg)Y\bigg(\frac{\cos(\chi'')}{\cos(\chi' - \chi'')}\bigg)}{{  \sqrt{\frac{H\cos^{2}( \chi'')}{\cos^{2}(\chi'-\chi'')}} \sqrt{\,E-\frac{H\cos^{2}( \chi'')}{\cos^{2}(\chi'-\chi'')}}}{  \sqrt{\frac{H\cos^{2}( \chi')}{\cos^{2}(\chi'-\chi'')}} \sqrt{\,\widetilde E-\frac{H\cos^{2}( \chi')}{\cos^{2}(\chi'-\chi'')}}}}   
\\
    &\times  \cos\left( \frac{2L'}{\hbar}\sqrt{\frac{E\cos^{2}(\chi'-\chi'')-H\cos^{2}( \chi'')}{H\cos^{2}( \chi'')}} -(m + m')\,\arccos\left(\sqrt{\frac{H}{E}}\frac{\cos(\chi'')}{\cos(\chi'-\chi'')}\right) \right)  
    \\&\times  \cos\left( \frac{2L''}{\hbar}\sqrt{\frac{\widetilde E\cos^{2}(\chi'-\chi'')-H\cos^{2}( \chi')}{H\cos^{2}( \chi')}} -(\widetilde{m} + \widetilde{m}')\,\arccos\left(\sqrt{\frac{H}{\widetilde E}}\frac{\cos(\chi')}{\cos(\chi'-\chi'')}\right) \right) 
\\ \tagg
    \label{B5}
    &\times \exp\bigg(\frac{2i}{\hbar}\bigg\{  -L\tan(\chi'-\chi'') - L'\tan(\chi'') + L''\tan(\chi')  \bigg\}\bigg).
\end{align*}

Observe that the following identity holds
    \begin{equation}
    \label{idcosexp}
    \int^\infty_{-\infty}\text{d}x \cos(ax - b)e^{itx} = \pi(e^{-ib}\delta(t + a) + e^{ib}\delta(t - a)).
\end{equation}
     
      Using Eq. (\ref{idcosexp}), we integrate over $L'$ and obtain 
\begin{align*}
     &\big( W_{Emm'}\star_{(\text{M})}W_{\widetilde{E}\widetilde{m}\widetilde{m}'}\big)(T,\,\chi,\,H,\,L) = \frac{N_{Emm'}N_{\widetilde{E}\widetilde{m}\widetilde{m}'}}{2\pi\hbar} e^{i(m-m'+\widetilde{m}-\widetilde{m}')\chi}\frac{1}{H}\int^{-\chi + 2\pi}_{-\chi}\text{d}\chi'
  \\
     &\times \int^{-\chi + 2\pi}_{-\chi}\text{d}\chi'' \int^\infty_{-\infty}\text{d}L'' \exp\bigg(\frac{2i}{\hbar}\bigg\{-L\tan(\chi'-\chi'') + L''\tan(\chi')\bigg\} \bigg)  e^{i(m - m')\chi'} e^{i(\widetilde{m} - \widetilde{m}')\chi''}
\\
 &\times Y\left(E-H\frac{\cos^{2}( \chi'')}{\cos^{2}(\chi'-\chi'')}\right)\, Y\left(\widetilde E-H\frac{\cos^{2}( \chi')}{\cos^{2}(\chi'-\chi'')}\right)  Y\bigg(\frac{\cos(\chi')}{\cos(\chi'-\chi'')}\bigg)Y\bigg(\frac{\cos(\chi'')}{\cos(\chi' - \chi'')}\bigg)
\\    
     &\times
     \frac{\cos^{2}(\chi'-\chi'')}{\cos(\chi')\cos(\chi'')}
     \frac{  1}{  \sqrt{\,E \cos^{2}(\chi'-\chi'') -H\cos^{2}( \chi'')}  \sqrt{\,\widetilde E \cos^{2}(\chi'-\chi'')- H\cos^{2}( \chi')}}
\\
    &\times  \cos\left( \frac{2L''}{\hbar}\sqrt{\frac{\widetilde E\cos^{2}(\chi'-\chi'')-H\cos^{2}( \chi')}{H\cos^{2}( \chi')}} -(\widetilde{m} + \widetilde{m}')\,\arccos\left(\sqrt{\frac{H}{\widetilde E}}\frac{\cos(\chi')}{\cos(\chi'-\chi'')}\right) \right) 
 \\
&\times  \exp\bigg(-i\,\text{sgn}(\tan(\chi''))(m + m')\,\arccos\bigg\{\sqrt{\frac{H}{E}}\frac{\cos(\chi'')}{\cos(\chi'-\chi'')}\bigg\}\bigg) 
\\ \tagg 
\label{B6}
&\times\delta\bigg(\text{sgn}(\tan(\chi'')) \tan(\chi'')-\sqrt{\frac{E\cos^{2}(\chi'-\chi'')-H\cos^{2}( \chi'')}{H\cos^{2}( \chi'')}}\bigg).
\end{align*}
The Dirac delta constraint in Eq. (\ref{B6}) is equivalent to the condition stating that
\begin{equation}
\label{eqchi1}
\cos^2(\chi'-\chi'') = \frac{H}{E}.
\end{equation}
Observe that formula (\ref{eqchi1}) treated as an equation   
with respect to $\chi'$ 
has solutions in the interval  $(-\chi,\,-\chi + 2\pi)$ only if $H < E$. To find them, for $H<E$ we build the following set:
\begin{equation}
        \mathcal{A} := \bigg\{\arccos\sqrt{\frac{H}{E}},\,\pi - \arccos\sqrt{\frac{H}{E}},\,\pi + \arccos\sqrt{\frac{H}{E}},\,2\pi - \arccos\sqrt{\frac{H}{E}}\bigg\}.
\end{equation}
There are four solutions of Eq. (\ref{eqchi1}) for $\chi'$
\begin{align}
\label{chi1-sol}
\chi'_0 = \chi'' + \alpha + 2n\pi,
\end{align}
where $\chi''\in(-\chi,\,-\chi+2\pi)$ and $\alpha\in\mathcal{A}$. The value of  $n\in\mathbb{Z}$ is chosen so that $\chi'_0 \in (-\chi,\,-\chi+2\pi)$. 

For the Dirac delta constraint 
\[F(\chi') := \text{sgn}(\tan(\chi'')) \tan(\chi'')-\sqrt{\frac{E\cos^{2}(\chi'-\chi'')-H\cos^{2}( \chi'')}{H\cos^{2}( \chi'')}}
\]
 we obtain
\begin{equation}
\label{chi1-det}
    \bigg|\frac{d F(\chi')}{d\chi'}\bigg|^{-1}_{\chi' = \chi'_0} = \frac{\sqrt{H} \sin ^2(\chi'')}{\sqrt{ \tan ^2(\chi'') (E-H)}} = \sqrt{\frac{H}{E - H}}\frac{|\cos(\chi'')|}{|\sin(\chi'')|}\sin^2(\chi'').
\end{equation}
Observe also that for every $\chi'_0$ the Heaviside function $Y\left(E-H\frac{\cos^{2}( \chi'')}{\cos^{2}(\chi'_0-\chi'')}\right) = 1$. 
Therefore, applying the identity (\ref{ident}) to the solutions (\ref{chi1-sol}) and applying the determinant (\ref{chi1-det}) we obtain
\begin{align*}
     &\big( W_{Emm'}\star_{(\text{M})}W_{\widetilde{E}\widetilde{m}\widetilde{m}'}\big)(T,\,\chi,\,H,\,L) = \frac{N_{Emm'}N_{\widetilde{E}\widetilde{m}\widetilde{m}'}}{2\pi\hbar} e^{i(m-m'+\widetilde{m}-\widetilde{m}')\chi}
\\
     &\times Y(E-H)\frac{1}{\sqrt{H(E - H)}}\sum_{\alpha\in\mathcal{A}} \int^{-\chi + 2\pi}_{-\chi}\text{d}\chi'' \int^\infty_{-\infty}\text{d}L''\frac{|\cos(\chi'')|}{|\sin(\chi'')|}\sin^2(\chi'')
   \\ &\times
     \exp\bigg(\frac{2i}{\hbar}\bigg\{-L\tan(\alpha) + L''\tan(\chi'' + \alpha)\bigg\} \bigg)  e^{i(m - m')(\chi''+\alpha)} e^{i(\widetilde{m} - \widetilde{m}')\chi''}
\\
 &\times  Y\left(\widetilde E-H\frac{\cos^{2}( \chi'' + \alpha)}{\cos^{2}(\alpha)}\right)  Y\bigg(\frac{\cos(\chi'' + \alpha)}{\cos(\alpha)}\bigg)Y\bigg(\frac{\cos(\chi'')}{\cos(\alpha)}\bigg)
\\    
     &\times
     \frac{\cos^{2}(\alpha)}{\cos(\chi'' + \alpha)\cos(\chi'')}
     \frac{  1}{   \sqrt{\,E \cos^{2}(\alpha) -H\cos^{2}( \chi'')}  \sqrt{\,\widetilde E \cos^{2}(\alpha)- H\cos^{2}( \chi''+\alpha)}}
\\
    &\times  \cos\left( \frac{2L''}{\hbar}\sqrt{\frac{\widetilde E\cos^{2}(\alpha)-H\cos^{2}( \chi''+\alpha)}{H\cos^{2}( \chi''+\alpha)}} -(\widetilde{m} + \widetilde{m}')\,\arccos\left(\sqrt{\frac{H}{\widetilde E}}\frac{\cos(\chi''+\alpha)}{\cos(\alpha)}\right) \right) 
\\ \tagg 
\label{B7}
&\times  \exp\bigg(-i\,\text{sgn}(\tan(\chi''))(m + m')\,\arccos\bigg\{\sqrt{\frac{H}{E}}\frac{\cos(\chi'')}{\cos(\alpha)}\bigg\}\bigg). 
\end{align*}

 Using the property  (\ref{idcosexp}) again, we integrate over $L''$ and obtain 
\begin{align*}
     &\big( W_{Emm'}\star_{(\text{M})}W_{\widetilde{E}\widetilde{m}\widetilde{m}'}\big)(T,\,\chi,\,H,\,L) = \frac{N_{Emm'}N_{\widetilde{E}\widetilde{m}\widetilde{m}'}}{4} e^{i(m-m'+\widetilde{m}-\widetilde{m}')\chi}
\\
     &\times Y(E-H)\frac{1}{\sqrt{H(E - H)}}\sum_{\alpha\in\mathcal{A}} \int^{-\chi + 2\pi}_{-\chi}\text{d}\chi'' \frac{|\cos(\chi'')|}{|\sin(\chi'')|}\sin^2(\chi'')
\\ &\times
     \exp\bigg(-\frac{2i}{\hbar}L\tan(\alpha) \bigg)  e^{i(m - m')(\chi''+\alpha)} e^{i(\widetilde{m} - \widetilde{m}')\chi''}
\\
 &\times  Y\left(\widetilde E-H\frac{\cos^{2}( \chi'' + \alpha)}{\cos^{2}(\alpha)}\right)  Y\bigg(\frac{\cos(\chi'' + \alpha)}{\cos(\alpha)}\bigg)Y\bigg(\frac{\cos(\chi'')}{\cos(\alpha)}\bigg)
\\  
     &\times
     \frac{\cos^{2}(\alpha)}{\cos(\chi'' + \alpha)\cos(\chi'')}
     \frac{  1}{  \sqrt{\,E \cos^{2}(\alpha) -H\cos^{2}( \chi'')}  \sqrt{\,\widetilde E \cos^{2}(\alpha)- H\cos^{2}( \chi''+\alpha)}}
\\
&\times  \exp\bigg(-i\,\text{sgn}(\tan(\chi''))(m + m')\,\arccos\bigg\{\sqrt{\frac{H}{E}}\frac{\cos(\chi'')}{\cos(\alpha)}\bigg\}\bigg)\\
&\times  \exp\bigg(i\,\text{sgn}(\tan(\chi'' + \alpha))(\widetilde{m} + \widetilde{m}')\,\arccos\bigg\{\sqrt{\frac{H}{\widetilde{E}}}\frac{\cos(\chi'' + \alpha)}{\cos(\alpha)}\bigg\}\bigg) 
\\ \tagg
\label{B8}
&\times\delta\bigg(\text{sgn}(\tan(\chi'' + \alpha)) \tan(\chi'' + \alpha)-\sqrt{\frac{\widetilde{E}\cos^{2}(\alpha)-H\cos^{2}( \chi'' + \alpha)}{H\cos^{2}( \chi'' + \alpha)}}\bigg).
\end{align*}
The Dirac delta constraint in Eq. (\ref{B8}) is independent of $\chi''$ and leads to the conclusion that $\widetilde{E} = E$.
Therefore
\[
\delta\bigg(\text{sgn}(\tan(\chi'' + \alpha)) \tan(\chi'' + \alpha)-\sqrt{\frac{\widetilde{E}\cos^{2}(\alpha)-H\cos^{2}( \chi'' + \alpha)}{H\cos^{2}( \chi'' + \alpha)}}\bigg) \sim \delta( \widetilde{E} - E).
\]
For the function
\[G(\widetilde{E}) := \text{sgn}(\tan(\chi'' + \alpha)) \tan(\chi'' + \alpha)-\sqrt{\frac{\widetilde{E}\cos^{2}(\alpha)-H\cos^{2}( \chi'' + \alpha)}{H\cos^{2}( \chi'' + \alpha)}}
\]
we calculate that
\begin{equation}
    \bigg|\frac{dG(\widetilde{E})}{d\widetilde{E}}\bigg|^{-1}_{\widetilde{E} = E} = 2E \,|\sin\big(\chi'' + \alpha \big)| |\cos\big(\chi'' + \alpha \big)|.
\end{equation}
Observe that $ Y\left(\widetilde E-H\frac{\cos^{2}( \chi'' + \alpha)}{\cos^{2}(\alpha)}\right) =1$ for $E = \widetilde{E}$.
Thus
\begin{align*}
  &    \big( W_{Emm'}\star_{(\text{M})}W_{\widetilde{E}\widetilde{m}\widetilde{m}'}\big)(T,\,\chi,\,H,\,L) = \frac{N_{Emm'}N_{\widetilde{E}\widetilde{m}\widetilde{m}'}}{2\sqrt{H(E - H)}}e^{i(m-m'+\widetilde{m}-\widetilde{m}')\chi} \delta(E - \widetilde{E})Y(E - H) \\ &\times  \sum_{\alpha\in \mathcal{A}}e^{i(m-m')\alpha}\exp\Big[-\frac{2i}{\hbar}L\tan(\alpha)\Big] \int^{-\chi+2\pi}_{-\chi}\text{d}\chi''  
   e^{i(m - m' + \widetilde{m} - \widetilde{m}')\chi''}  \\
    &\times \exp\Big(i\,\text{sgn}(\tan(\chi'' + \alpha))\,(\widetilde{m} + \widetilde{m}')\,\arccos\{|\cos(\chi''+\alpha)|\}\Big) \\ \tagg
     &\times 
  \exp\Big(-i\,\text{sgn}(\tan( \chi''))\,(m + m')\,\arccos\{|\cos(\chi'')|\}\Big)  Y\bigg(\frac{\cos( \chi''+\alpha)}{\cos(\alpha)}\bigg)Y\bigg(\frac{\cos(\chi'')}{\cos(\alpha)}\bigg).
\end{align*}

Notice that for $\theta \neq (k + \frac{1}{2}) \pi, \;\; k \in {\mathbb Z}$
  \begin{equation}
  \label{identity2}
\text{sgn}(\tan\theta)\,\arccos|\cos\theta|
=\theta-\pi\,\Big\lfloor\frac{\theta+\frac{\pi}{2}}{\pi}\Big\rfloor,
\end{equation}
where $\lfloor \cdot \rfloor$ denotes the floor function. Hence, after changing the variable $\chi'' \rightarrow \chi''- \chi$, we obtain
\begin{align*}
    &    \big( W_{Emm'}\star_{(\text{M})}W_{\widetilde{E}\widetilde{m}\widetilde{m}'}\big)(T,\,\chi,\,H,\,L) = \frac{N_{Emm'}N_{\widetilde{E}\widetilde{m}\widetilde{m}'}}{2\sqrt{H(E - H)}} \delta(E - \widetilde{E})Y(E - H) 
   e^{i(m + m'-\widetilde{m} - \widetilde{m}' )\chi} 
    \\ 
    &\times \sum_{\alpha\in \mathcal{A}} e^{i(m - m'+\widetilde{m} + \widetilde{m}' )\alpha} \exp\Big[-\frac{2i}{\hbar}L\tan(\alpha)\Big] 
    \int^{2\pi}_{0} \text{d}\chi''e^{2i(\widetilde{m} - m' )\chi''}  \\ 
    &\times \underbrace{\exp\bigg(-i(\widetilde{m} + \widetilde{m}')
   \pi \Big\lfloor \frac{\chi''-\chi +\alpha + \frac{\pi}{2}}{\pi}  \Big\rfloor\bigg) 
    \exp\bigg(i(m + m') \big( \pi \Big\lfloor \frac{\chi''-\chi + \frac{\pi}{2}}{\pi}  \Big\rfloor\bigg)}_{=:\mathcal{C}_\alpha}
     \\ \tagg \label{intYY} &\times Y\bigg(\frac{\cos( \chi''-\chi+\alpha)}{\cos(\alpha)}\bigg)Y\bigg(\frac{\cos(\chi''-\chi)}{\cos(\alpha)}\bigg).
\end{align*}

In order to determine for which ranges of $\chi''-\chi$ the product $Y\bigg(\frac{\cos( \chi''-\chi+\alpha)}{\cos(\alpha)}\bigg)Y\bigg(\frac{\cos(\chi''-\chi)}{\cos(\alpha)}\bigg)$ equals $1$, we have to examine four cases with respect to $\alpha$. We present the conclusions in Table \ref{tabela}. 
\begin{table}[H]
    \centering
    \begin{tabular}{c|c|c|c|c|c}
      $\alpha$ & $u_-$ & $u_+$ & $\Big\lfloor \frac{\chi''-\chi +\alpha + \frac{\pi}{2}}{\pi}  \Big\rfloor$ &  $\Big\lfloor \frac{\chi''-\chi  + \frac{\pi}{2}}{\pi}  \Big\rfloor$ & $\mathcal{C}_\alpha$  \\ \hline
     $\big(0,\,\frac{\pi}{2}\big)$    & $-\frac{\pi}{2}$ & $\frac{\pi}{2}-\alpha$ & $0$ & $0$ & $1$ \\
       $\big(\frac{\pi}{2},\,\pi\big)$    & $\frac{\pi}{2}$ & $\frac{3\pi}{2}-\alpha$ & $1$ & $1$ & $e^{i\pi (-\widetilde{m}-\widetilde{m}'+m+m')}$ \\
         $\big(\pi,\,\frac{3\pi}{2}\big)$    & $\frac{5\pi}{2}-\alpha$ & $\frac{3\pi}{2}$ & $1$ & $3$ & $e^{i\pi (-3\widetilde{m}-3\widetilde{m}'+m+m')}$ \\
           $\big(\frac{3\pi}{2},\,2\pi\big)$    & $\frac{3\pi}{2}-\alpha$ & $\frac{\pi}{2}$ & $0$ & $2$ & $e^{i\pi (-2\widetilde{m}-2\widetilde{m}')}$
    \end{tabular}
    \caption{Four ranges of $\alpha$ for which if $(\chi''-\chi)\in (u_-,\,u_+)$ then $Y\bigg(\frac{\cos( \chi''-\chi+\alpha)}{\cos(\alpha)}\bigg)Y\bigg(\frac{\cos(\chi''-\chi)}{\cos(\alpha)}\bigg)=1$. Constant $\mathcal{C}_\alpha$ is defined in Eq. (\ref{intYY}).}
    \label{tabela}
\end{table}
From Table \ref{tabela}, it can be concluded that we can factor exponents with  $\Big\lfloor \frac{\chi''-\chi +\alpha + \frac{\pi}{2}}{\pi}  \Big\rfloor$ and $\Big\lfloor \frac{\chi''-\chi  + \frac{\pi}{2}}{\pi}  \Big\rfloor$ out of the integral. Using the indefinite integrals $\int \text{d}x e^{ia x} =\frac{1}{a} e^{iax}$ if $a\neq 0$ and $\int \text{d}x e^{ia x} =x$ if $a = 0$ and information from the table, one can obtain
\begin{align*}
    &    \big( W_{Emm'}\star_{(\text{M})}W_{\widetilde{E}\widetilde{m}\widetilde{m}'}\big)(T,\,\chi,\,H,\,L) = \frac{N_{Emm'}N_{\widetilde{E}\widetilde{m}\widetilde{m}'}}{2\sqrt{H(E - H)}} \delta(E - \widetilde{E})Y(E - H) 
   e^{i(m + m'-\widetilde{m} - \widetilde{m}' )\chi} 
    \\ \tagg \label{WWf}
    &\times \sum_{\alpha\in \mathcal{A}} e^{i(m - m'+\widetilde{m} + \widetilde{m}' )\alpha} \exp\Big[-\frac{2i}{\hbar}L\tan(\alpha)\Big] 
   \mathcal{C}_\alpha \mathcal{F}_\alpha ,
\end{align*}
where for each $\alpha \in \mathcal{A}$, $A := (\chi+u_-)\text{mod}\,2\pi$ and $B := (\chi+u_+)\text{mod}\,2\pi$
\begin{align}
    \mathcal{F}_\alpha := \begin{cases}
        |\pi-\alpha| &, \, \widetilde{m} = m', \\
        \frac{1}{2i (\widetilde{m}-m')}\big[e^{2i(\widetilde{m}-m')B} - e^{2i(\widetilde{m}-m')A} + Y(A-B)(e^{i4\pi(\widetilde{m}-m')} - 1)  \big] &, \widetilde{m} \neq m'.
    \end{cases} 
\end{align}

If we assume that $m,\,m',\,\widetilde{m},\,\widetilde{m}'\in\mathbb{Z}$ and use  the trigonometric identity $\tan \big( \arccos(x)\big)= \frac{\sqrt{1-x^2}}{x}$, then Eq. (\ref{WWf}) reduces to the form
\begin{align*}
   &    \big( W_{Emm'}\star_{(\text{M})}W_{\widetilde{E}\widetilde{m}\widetilde{m}'}\big)(T,\,\chi,\,H,\,L) = \pi N_{Emm'}N_{\widetilde{E}\widetilde{m}\widetilde{m}'} \frac{1}{\sqrt{H(E - H)}}\delta(E - \widetilde{E}) Y(E - H)\\ &\times e^{i(m - \widetilde{m}')\chi}\cos\left(\frac{2L}{\hbar}\sqrt{\frac{E-H}{H}}-(m + \widetilde{m}')\,\arccos\sqrt{\frac{H}{E}}\right)\delta_{m'\widetilde{m}} \\
   \label{norm1}\tagg &= \frac{\pi N_{Emm'}N_{\widetilde{E}\widetilde{m}\widetilde{m}'}}{N_{Em\widetilde{m}'}} \delta(E - \widetilde{E}) W_{Em\widetilde{m}'}(T,\,\chi,\,H,\,L)\delta_{m'\widetilde{m}}.
\end{align*}
In addition, if $m=m'$ and $\widetilde{m}=\widetilde{m}'$, then Eq. (\ref{norm1}) becomes
\begin{align}
\label{WEmm1B_final}
(W_{Em}\star_{(\text{M})}W_{\widetilde{E}\widetilde{m}}\big)(T,\,\chi,\,H,\,L) =\pi N_{Em}\delta(E - \widetilde{E}) W_{Em}(T,\,\chi,\,H,\,L) \delta_{m\widetilde{m}}.
\end{align}

If at least one of the numbers $m,\,m',\,\widetilde{m},\,\widetilde{m}'$ belongs to $\mathbb{R}\setminus\mathbb{Z}$, then, in Eq. (\ref{WWf}), we cannot simplify the coefficients $\mathcal{C}_\alpha$.

\renewcommand{\thesection}{\Alph{section}}
\renewcommand{\theequation} {\Alph{section}.\arabic{equation}}
\setcounter{section}{2}
\setcounter{equation}{0}
\section{Calculation of the Moyal product of Wigner functions  $(W_{Emm'}\star_{(\text{M})}W_{\widetilde{p}_{x0}\widetilde{p}_{y0}})(T,\,\chi,\,H,\,L)$}
\label{appendixC}
The Moyal product of functions  $(W_{Emm'}\star_{(\text{M})}W_{\widetilde{p}_{x0}\widetilde{p}_{y0}})(T,\,\chi,\,H,\,L)$ is given by the integral
\begin{align*}
    &\big( W_{Emm'}\star_{(\text{M})}W_{\widetilde{p}_{x0}\widetilde{p}_{y0}}\big)(T,\,\chi,\,H,\,L) = \frac{N_{Emm'}N_{\widetilde{p}_{x0}\widetilde{p}_{y0}}}{(\pi \hbar)^4} \\
     &\times \int^\infty_{-\infty}\text{d}T'\int^{ 2\pi}_{0}\text{d}\chi'\int^\infty_0 \text{d}H' \int^\infty_{-\infty}\text{d}L' \int^\infty_{-\infty}\text{d}T''\int^{ 2\pi}_{0}\text{d}\chi''\int^\infty_0 \text{d}H'' \int^\infty_{-\infty}\text{d}L''\\  
     &\times Y(E - H')\frac{1}{\sqrt{H'(E - H')}}  e^{i(m - m')\chi'} 
\cos\bigg(\frac{2 L'}{\hbar}\sqrt{\frac{E - H'}{H'}} - (m + m')\arccos\sqrt{\frac{H'}{E}}\bigg) \\ 
&\times \delta(H'' - \widetilde{E})\delta(\chi'' - \widetilde{\chi}_0)  \exp\bigg(\frac{2i}{\hbar}\bigg\{ 
 2T\bigg[\sqrt{H H'}\cos(\chi - \chi') -\sqrt{HH''}\cos(\chi - \chi'')  \bigg] \\ &+ 2T' \bigg[\sqrt{H'H''}\cos(\chi' - \chi'') - \sqrt{HH'}\cos(\chi - \chi')\bigg]  \\ &+ 2T''\bigg[\sqrt{HH''}\cos(\chi - \chi'') - \sqrt{H'H''}\cos(\chi'-\chi'')\bigg] \\ &+ 
  L \bigg[\sqrt{\frac{H'}{H}}\sin(\chi - \chi') - \sqrt{\frac{H''}{H}}\sin(\chi - \chi'')\bigg] \\ &+ L'\bigg[\sqrt{\frac{H}{H'}}\sin(\chi - \chi') + \sqrt{\frac{H''}{H'}}\sin(\chi' - \chi'')\bigg] \\ \tagg  &+ L''\bigg[\sqrt{\frac{H'}{H''}}\sin(\chi'-\chi'') - \sqrt{\frac{H}{H''}}\sin(\chi - \chi'')\bigg] \bigg\}\bigg).
\end{align*} 
We integrate over $H''$ and $\chi''$ first.
\begin{align*}
    &\big( W_{Emm'}\star_{(\text{M})}W_{\widetilde{p}_{x0}\widetilde{p}_{y0}}\big)(T,\,\chi,\,H,\,L) = \frac{N_{Emm'}N_{\widetilde{p}_{x0}\widetilde{p}_{y0}}}{(\pi \hbar)^4} \\
     &\times \int^\infty_{-\infty}\text{d}T'\int^{ 2\pi}_{0}\text{d}\chi'\int^\infty_0 \text{d}H' \int^\infty_{-\infty}\text{d}L' \int^\infty_{-\infty}\text{d}T''
      \int^\infty_{-\infty}\text{d}L''\\  
     &\times Y(E - H')\frac{1}{\sqrt{H'(E - H')}}  e^{i(m - m')\chi'} 
\cos\bigg(\frac{2 L'}{\hbar}\sqrt{\frac{E - H'}{H'}} - (m + m')\arccos\sqrt{\frac{H'}{E}}\bigg) \\ 
&\times \exp\bigg(\frac{2i}{\hbar}\bigg\{ 
 2T\bigg[\sqrt{H H'}\cos(\chi - \chi') -\sqrt{H\widetilde{E}}\cos(\chi - \widetilde{\chi}_0)  \bigg] \\ 
 &+ 2T' \bigg[\sqrt{H'\widetilde{E}}\cos(\chi' - \widetilde{\chi}_0) - \sqrt{HH'}\cos(\chi - \chi')\bigg]  \\
  &+ 2T''\bigg[\sqrt{H\widetilde{E}}\cos(\chi - \widetilde{\chi}_0) - \sqrt{H'\widetilde{E}}\cos(\chi'- \widetilde{\chi}_0)\bigg] \\ &+ 
  L \bigg[\sqrt{\frac{H'}{H}}\sin(\chi - \chi') - \sqrt{\frac{\widetilde{E}}{H}}\sin(\chi - \widetilde{\chi}_0)\bigg] \\ &+ L'\bigg[\sqrt{\frac{H}{H'}}\sin(\chi - \chi') + \sqrt{\frac{\widetilde{E}}{H'}}\sin(\chi' - \widetilde{\chi}_0)\bigg] \\ \tagg  &+ L''\bigg[\sqrt{\frac{H'}{\widetilde{E}}}\sin(\chi'- \widetilde{\chi}_0) - \sqrt{\frac{H}{\widetilde{E}}}\sin(\chi - \widetilde{\chi}_0)\bigg] \bigg\}\bigg).
\end{align*} 

In the next step, we integrate over $T' $ , $T''$ and $L''$ using the identities  $\int^\infty_{-\infty}\text{d}x e^{ikx} = 2\pi\delta(k)$ and $\delta(a x) = \delta(x)/|a|.$ Thus 
\begin{align*}
    &\big( W_{Emm'}\star_{(\text{M})}W_{\widetilde{p}_{x0}\widetilde{p}_{y0}}\big)(T,\,\chi,\,H,\,L) = \frac{N_{Emm'}N_{\widetilde{p}_{x0}\widetilde{p}_{y0}}}{4\pi \hbar}  \frac{1}{\sqrt{H'}}\int^{ 2\pi}_{0}\text{d}\chi'\int^\infty_0 \text{d}H' \int^\infty_{-\infty}\text{d}L' 
      \\  
     &\times Y(E - H')\frac{1}{\sqrt{H'(E - H')}} e^{i(m - m')\chi'} 
\cos\bigg(\frac{2 L'}{\hbar}\sqrt{\frac{E - H'}{H'}} - (m + m')\arccos\sqrt{\frac{H'}{E}}\bigg) \\ 
& \times \delta \bigg[\sqrt{\widetilde{E}}\cos(\chi' - \widetilde{\chi}_0) - \sqrt{H}\cos(\chi - \chi')\bigg]
\\
& \times \delta \bigg[\sqrt{H}\cos(\chi - \widetilde{\chi}_0) - \sqrt{H'}\cos(\chi'- \widetilde{\chi}_0)\bigg]                                        
\\
& \times \delta \bigg[\sqrt{H'} \sin(\chi' - \widetilde{\chi}_0)
-\sqrt{H} \sin(\chi - \widetilde{\chi}_0)
\bigg] \\
&\times \exp\bigg(\frac{2i}{\hbar}\bigg\{ 
 2T\bigg[\sqrt{H H'}\cos(\chi - \chi') -\sqrt{H\widetilde{E}}\cos(\chi - \widetilde{\chi}_0)  \bigg]  \\ &+ 
  L \bigg[\sqrt{\frac{H'}{H}}\sin(\chi - \chi') - \sqrt{\frac{\widetilde{E}}{H}}\sin(\chi - \widetilde{\chi}_0)\bigg] \\ &+ L'\bigg[\sqrt{\frac{H}{H'}}\sin(\chi - \chi') + \sqrt{\frac{\widetilde{E}}{H'}}\sin(\chi' - \widetilde{\chi}_0)\bigg].
\end{align*} 

The last two Dirac deltas imply that
$
H'=H $ and $  \chi'=\chi
$. Hence, the first Dirac delta constraint has the solution $\widetilde{E} = \frac{H}{\cos^2(\chi-\widetilde{\chi}_0)}$ that exists only if $\cos(\chi - \widetilde{\chi}_0)>0$.

Thus, our product of the Dirac deltas is equal to
\[
\frac{4 \sqrt{H }}{\cos^2(\chi -\widetilde{\chi}_0 )}
Y\big(\cos(\chi - \widetilde{\chi}_0) \big)
 \delta(\chi'-\chi)\delta(H'-H)\delta \bigg(\widetilde{E} - \frac{H}{\cos^2(\chi - \widetilde{\chi}_0)} \bigg),
\]
and the calculated product simplifies to the form
\begin{align*}
    &\big( W_{Emm'}\star_{(\text{M})}W_{\widetilde{p}_{x0}\widetilde{p}_{y0}}\big)(T,\,\chi,\,H,\,L) = \frac{N_{Emm'}N_{\widetilde{p}_{x0}\widetilde{p}_{y0}}}{\pi \hbar}
    \frac{ Y\big(\cos(\chi - \widetilde{\chi}_0) \big)  }{\cos^2(\chi -\widetilde{\chi}_0 )}
    \delta \bigg(\widetilde{E} - \frac{H}{\cos^2(\chi - \widetilde{\chi}_0)} \bigg)
    \\&
     \times Y(E - H)\frac{1}{\sqrt{H(E - H)}}  e^{i(m - m')\chi} 
     \exp\bigg(-\frac{2i}{\hbar}L \tan(\chi -\widetilde{\chi}_0)\bigg)
      \\  
     & 
     \times \int^\infty_{-\infty}\text{d}L' 
\cos\bigg(\frac{2 L'}{\hbar}\sqrt{\frac{E - H}{H}} - (m + m')\arccos\sqrt{\frac{H}{E}}\bigg) \exp\bigg(\frac{2i}{\hbar}  L' \tan(\chi - \widetilde{\chi}_0)\bigg).
\end{align*} 
From the identity \eqref{idcosexp} we find that the integral with respect to $L'$ equals
\[
\frac{\pi \hbar}{2} \bigg[ \exp\bigg( -i(m+m') \arccos\sqrt{\frac{H}{E}} \bigg) \bigg]
\delta \bigg( \tan (\chi - \widetilde{\chi}_0 ) + 
\sqrt{\frac{E-H}{H}}\bigg)
\]
\[
+\frac{\pi\hbar}{2} \bigg[ \exp\bigg( i(m+m') \arccos\sqrt{\frac{H}{E}} \bigg) \bigg]
\delta \bigg( \tan (\chi - \widetilde{\chi}_0 ) - 
\sqrt{\frac{E-H}{H}}\bigg).
\]
Only one of these terms survives. If $\sin(\chi - \widetilde{\chi}_0 )>0$ then the second term in the sum contributes. If $\sin(\chi - \widetilde{\chi}_0 )<0,$ then the first term contributes. However, both cases lead to the same conclusion: $E= \frac{H}{\cos^2(\chi - \widetilde{\chi}_0)} =\widetilde{E}.$
Moreover, if an angle $\alpha \in (\frac{3\pi}{2},2 \pi)$, then $\arccos (\cos(\alpha))=2 \pi - \alpha$.

 Therefore,
\begin{align*}
    &\big( W_{Emm'}\star_{(\text{M})}W_{\widetilde{p}_{x0}\widetilde{p}_{y0}}\big)(T,\,\chi,\,H,\,L) = N_{Emm'}N_{\widetilde{p}_{x0}\widetilde{p}_{y0}}
    \frac{ Y\big(\cos(\chi - \widetilde{\chi}_0) \big)  }{\cos^2(\chi -\widetilde{\chi}_0 )}  e^{2im \chi}   e^{-i(m + m')\widetilde{\chi}_0 }
    \\ 
    \tagg \label{WEmm1Wpx0py0_finalC}
    &\times 
     \exp\bigg(-\frac{2i}{\hbar}L \tan(\chi -\widetilde{\chi}_0)\bigg) \delta \bigg(E-\widetilde{E}\bigg)
     \delta \bigg(E - \frac{H}{\cos^2(\chi - \widetilde{\chi}_0)} \bigg).
\end{align*}


\begin{thebibliography}{99}

\bibitem{YK91}
Y. S. Kim and M. E. Noz, {\it Phase Space Picture of Quantum Mechanics}, World Scientific, Singapore (1991).

\bibitem{FS94}
F. E. Schroeck, Jr. , {\it Quantum Mechanics on Phase Space}, Kluwer Academic Publishers, Dordrecht (1996).

\bibitem{WS01} W. Schleich, {\it Quantum Optics in Phase Space}, Wiley-VCH, Berlin (2001).

\bibitem{CZ05} C. K. Zachos, D. B. Fairlie and T. L. Curtright (Eds.), {\it Quantum Mechanics in Phase Space}, World Scientific, London (2005).

\bibitem{tat}
W. I. Tatarskij, {\it Usp. Fiz. Nauk} {\bf 139}, 587 (1983) (in Russian).

\bibitem{cd}
M. Hillery, R.F. O'Connell, M.O. Scully and E.P. Wigner, {\it
Phys. Rep.} {\bf 106}, 121 (1984).

\bibitem{lee}
H. W. Lee, {\it Phys. Rep.} {\bf 259}, 147 (1995).

\bibitem{dit}
G. Dito and D. Sternheimer, {\it Deformation Quantisation: Genesis, Developments and Metamorphoses}, in {\it Deformation quantisation}, ed. G. Halbout, IRMA Lectures Maths. Theor. Phys., Walter de Gruyter, Berlin 2002, p. 9. 

\bibitem{SW07} S. Waldmann, {\it Poisson-Geometrie und Deformationsquantisierung}, Springer, Berlin 2007 (in German).

 \bibitem{IS1} R Simon, G. S. Agarwal,  \textit{Opt Lett.} \textbf{25}, 1313 (2000).

\bibitem{IS2}  J. P. Dahl, W. P. Schleich, \textit{Phys. Rev. A} \textbf{79}, 024101 (2009). 

\bibitem{IS3} J. P. Dahl, M. Springborg \textit{Mol. Phys.}, \textbf{47}, 1001 (1982).
\bibitem{IS4} L. Praxmeyer, J. Mostowski and K. Wódkiewicz, \textit{J. Phys. A: Math Gen.} \textbf{39}, 14143 (2006).

\bibitem{IS5} P. Campos, M. G. R. Martins, M. C. B. Fernandes, J. D. M. Vianna, {\it Ann. Phys.} \textbf{390}, 60 (2018), \url{https://doi.org/10.1016/j.aop.2018.01.002}.

\bibitem{IS6} M. Kordell II, R. J. Fries, C. M. Ko, {\it Ann. Phys.} \textbf{443}, 168960 (2022).


    \bibitem{Fedosov1}
B.~V.~Fedosov, {\it J. Diff. Geom.} \textbf{40}, 213 (1994).
\bibitem{Fedosov2} B.~V.~Fedosov, {\it Deformation quantisation and Index Theory}, Wiley (1995).

\bibitem{F1} J. Tosiek and M. Przanowski, {\it Entropy} \textbf{23}, 581 (2021).
\bibitem{F2} J. Tosiek, {\it J. Phys.: Conf. Ser.} \textbf{128}, 012024 (2008).
\bibitem{F3} M. Bordemann, N. Neumaier and S. Waldmann, {\it Comm. in Math. Phys.} \textbf{198}, 363 (1998).

\bibitem{Tosiek_oscillator} J.~Tosiek, {\it Phys. Lett. A} \textbf{376}, 2023 (2012).

\bibitem{Schleich0}
 J.~P.~Dahl, S.~Varró, A.~Wolf and W.~P.~Schleich,
    {\it J. of Mod. Opt.} \textbf{54}, 2017 (2007).

 \bibitem{Schleich}
    W.~P.~Schleich, J.~P.~Dahl and S.~Varró,
    {\it Opt. Comm.} \textbf{283}, 786 (2010).

\bibitem{Arnold} W. I. Arnold, \textit{Mathematical methods of classical mechanics 2nd ed.}, Springer-Verlag (1989).

\bibitem{Libermann} P. Libermann and C--M. Marle, \textit{Symplectic Geometry and Analytical Mechanics}, D. Reidel Publishing Company (1987).



\bibitem{funkcje} J.~M.~Gracia-Bondía and J.~C.~Várilly, {\it J. Phys. A: Math. Gen.} \textbf{21}, 879 (1988).
\bibitem{Przanowski}
 J.~F.~Plebański, M.~Przanowski and J.~Tosiek,  {\it Acta Phys. Pol.} \textbf{B 27}, 1961 (1996).

\bibitem{cont1} J. G. Muga, R. Sala and S. Brouard, \textit{Sol. St. Comm.} \textbf{94}, 877 (1995).

\bibitem{cont2} B. Belchev and M. A. Walton, arXiv.org:1001.4816v1, (2010).

\bibitem{cont3} D. M. Heim, W. P. Schleich, P. M. Alsing, J. P. Dahl, S. Varro, \textit{Phys. Lett. A} \textbf{377}, 1822 (2013).

\bibitem{cont4} K. B. Wolf, M. A. Alonso and G. W. Forbes, \textit{J. Opt. Soc. Am. A} \textbf{16}, 2476 (1999).





\bibitem{symplectic} I.~Gelfand, V.~Retakh and M.~Shubin, {\it Adv. Math} \textbf{136}, 104 (1998).

\bibitem{gadella1} M.~Gadella, M.~A.~del Olmo and J.~Tosiek, {\it Ann. Phys.}
 \textbf{307}, 272 (2003).
 
 \bibitem{Gadella} M.~Gadella, M.~A.~del Olmo and J.~Tosiek, {\it J. Geom. Phys.} \textbf{55}, 316 (2005).


\bibitem{semiclassical}
M. Sirugue, M. Sirugue-Collin and A. Truman, {\it Ann. de l'I.H.P. Phys. théor.} \textbf{41}, 429 (1984).

\bibitem{polar} W.~E.~Weisstein, {\it Jacobi-Anger expansion}. 
From MathWorld-A Wolfram Resource. \url{https://mathworld.wolfram.com/Jacobi-AngerExpansion.html}, access: 15.01.2026.

\bibitem{Varro} J.~P.~Dahl, S.~Varró, A.~Wolf and W.~P.~Schleich, {\it Phys. Rev. A} \textbf{75}, 052107 (2007).

\bibitem{hydrogen} L.~G.~Mardoyan, G.~S.~Pogosyan, A.~N.~Sissakian and V.~M~Ter-Antonyan, {\it Quant. Sys. with Hidd. Symm.
Int. Exp.}, e-Print: 2310.17336 [math-ph] (2023).

\bibitem{Schwartz}
    L.~Schwartz,
    {\it Mathematics for the Physical Sciences}, Hermann, Addison-Wesley (1966).




\end{thebibliography}
\end{document}